\documentclass[useAMS,usenatbib]{mnras}
\usepackage{lmodern}
\usepackage{newtxtext,newtxmath}

\usepackage[T1]{fontenc}

\DeclareRobustCommand{\VAN}[3]{#2}
\let\VANthebibliography\thebibliography
\def\thebibliography{\DeclareRobustCommand{\VAN}[3]{##3}\VANthebibliography}

\usepackage{graphicx}	
\usepackage{amsmath}	

\title[\textsc{Hi} survey of the Shapley Supercluster]{Neutral hydrogen in the Shapley Supercluster core -- II. Gas content\\ and star formation in the filaments connecting the core}

\author[L. Gwebushe et al.]{
L. Gwebushe,$^{1,2}$\thanks{E-mail: nobonke123@gmail.com} T. Venturi,$^{2,1}$ P. Merluzzi,$^{3}$ G. Busarello,$^{3}$ O. Smirnov,$^{1,4,2,5,6}$ V. Casasola,$^{2}$ \newauthor F. La Barbera,$^{3}$ M. Ramatsoku,$^{1,7}$ and J. Dawson,$^{1}$ 
\\ \\
$^{1}$ Centre for Radio Astronomy Techniques and Technologies (RATT), Department of Physics and Electronics, Rhodes University, Makhanda 6140,\\ South Africa\\
$^{2}$ INAF–Istituto di Radioastronomia, via Gobetti 101, I–40129, Bologna, Italy\\
$^{3}$ INAF–Osservatorio Astronomico di Capodimonte, Via Moiariello 16, I–80131 Napoli, Italy\\
$^{4}$ South African Radio Astronomy Observatory, 2 Fir Street, Black River Park, Observatory, Cape Town 7925, South Africa\\
$^{5}$ Astrophysics, Department of Physics, University of Oxford, Keble Road, Oxford, OXI 3RH, UK\\
$^{6}$ Breakthrough Listen, Astrophysics, Department of Physics, University of Oxford, Keble Road, Oxford, OXI 3RH, UK\\
$^{7}$ INAF–Osservatorio Astronomico di Cagliari, Via della Scienza 5, I-09047 Selargius, (CA), Italy
}
\date{Accepted 2026 November 18. Received 2026 November 17; in original form 2026 August 17}

\pubyear{\the\year{}}

\begin{document}
\label{firstpage}
\pagerange{\pageref{firstpage}--\pageref{lastpage}}
\maketitle

\begin{abstract}
We investigate the neutral atomic hydrogen (\textsc{Hi}) content of galaxies across the Shapley Supercluster (SSC) at $\langle z \rangle \sim 0.048$ by combining new deep MeerKAT L-band observations of five filament pointings with archival MeerKAT Galaxy Cluster Legacy Survey observations of the supercluster core. Together with optical data from the Shapley Supercluster Survey, these observations provide a homogeneous view of galaxy evolution from the virialised cluster environment to the surrounding filaments. We compare the \textsc{Hi} content, star formation activity, gas depletion timescales, and morphological properties of galaxies relative to the gas fraction and star-forming main sequences. The global \textsc{Hi} detection fraction increases from $\sim$30 per cent in the cluster centre to $\sim 45$ per cent in the filaments. While \textsc{Hi}-detected galaxies in the cluster centre are dominated by transition-zone systems, the filament population is shared nearly equally between star-forming main sequence and transition-zone detections, indicating ongoing environmental transformation. Their gas properties also differ: cluster galaxies exhibit progressively depleted \textsc{Hi} reservoirs towards the red sequence, whereas filament galaxies remain primarily gas-rich. Within the filaments, long \textsc{Hi} depletion timescales exceeding the Hubble time are concentrated in specific populations, notably $\sim71$ per cent of transition-zone and $\sim57$ per cent of gas-rich detections, indicating the inefficient conversion of atomic gas into stars. The \textsc{Hi}-detected filament population is dominated by late-type galaxies, while the underlying non-detected population is primarily composed of early-type systems, consistent with environmental pre-processing before cluster infall. 
\end{abstract}

\begin{keywords}
galaxies: evolution -- large-scale structure of Universe -- galaxies: clusters: individual: Abell\,3558 -- galaxies: clusters: individual: SC\,1329-313 -- galaxies: star formation -- radio lines: galaxies -- intergalactic medium\end{keywords}



\section{Introduction}
\label{sec:Introduction}

Superclusters are the most massive gravitationally bound structures in the Universe ($M\sim10^{16}$~M$_{\odot}$). They form within the supercluster cocoons, or basins of attraction, of the Cosmic Web and grow through the hierarchical gravitational collapse of primordial density perturbations \citep{einasto2021evolution}. Superclusters contain structures spanning a wide range of masses, from the most massive galaxy clusters  ($M\sim10^{15}$~M$_{\odot}$) to small groups ($M\sim10^{13}$~M$_{\odot}$), as well as the lower-density filamentary regions that connect them \citep[e.g][]{voit2005tracing,kravtsov2012formation,cautun2014evolution,libeskind2018tracing}. During their assembly, superclusters experience a variety of dynamical processes, ranging from relatively rare major mergers involving intermediate- to massive clusters (i.e. $\ge 6\times10^{14}$~M$_{\odot}$) with mass ratios $M_2/M_1\geq$\,1:4--1:5, to the more frequent accretion of systems through minor mergers involving either lower masses or significantly smaller mass ratios \citep[e.g.][]{cassano2016can}. These processes affect all cluster constituents, including the intra-cluster medium \citep{kravtsov2012formation}, the diffuse radio emission permeating large volumes of the cluster environment \citep[see][for an observational review]{van2019diffuse}, and the galaxies residing within them.

Due to the broad range of galaxy densities and dynamical states they encompass, superclusters provide ideal laboratories for a variety of studies. These include testing predictions of hierarchical mass assembly, investigating dynamical processes such as cluster mergers and group accretion, and examining the role of environment in galaxy evolution across a continuous range of densities, from massive cluster cores to filaments, at a common redshift.

The target of this paper is the core of the Shapley Supercluster \citep{shapley1930note}, hereafter SSC, the most massive gravitationally bound collapsing supercluster in the local Universe. The SSC contains at least 25 Abell clusters and tens of smaller groups within the redshift range $0.03<z<0.06$, distributed over an area of 150\,deg$^2$ in the Southern Hemisphere \citep{raychaudhury1989distribution,
einasto2001optical,de2005measuring,quintana2020redshift}. These structures are connected by a complex network of galaxy filaments \citep{quintana2000shapley,proust2006structure,haines2018shapley,quintana2020redshift}. The densest region of the SSC, located at $\langle z \rangle = 0.048$, comprises three Abell clusters and two poor clusters arranged in a continuous filamentary structure extending over 2 degrees ($\sim$7 Mpc) and permeated by hot gas \citep{collaboration2014planck,merluzzi2015shapley}. This region is the most strongly affected by the ongoing large-scale collapse and has been extensively studied at radio wavelengths, where merger and accretion processes leave clear signatures from inter-cluster scales down to individual galaxies \citep[e.g.][]{venturi2022radio,giacintucci2022candle,merluzzi2024ram,trehaeven2025peculiar,trehaeven2026direction,ignesti2026supersonic}.

In particular, two clusters in the SSC, A\,3558 and A\,3562, host diffuse radio emission on cluster scales \citep[a mini-halo and a radio halo, respectively; see][]{venturi2022radio,giacintucci2022candle,trehaeven2025peculiar}. This emission is interpreted as the result of turbulent re-acceleration operating within the intra-cluster medium and driven by minor merger activity \citep[see][for a theoretical review]{brunetti2014cosmic}. In addition, a Mpc-scale bridge of extremely low-surface-brightness diffuse radio emission has been detected with MeerKAT and ASKAP between A\,3562 and the group SC\,1329--313. This feature has been interpreted as evidence of turbulence generated during the dynamical interaction between the two systems \citep{finoguenov2004xmm,venturi2022radio}. The same deep MeerKAT and ASKAP observations, complemented by uGMRT data, have also revealed several tailed radio sources associated with star-forming galaxies. These have been interpreted as signatures of merger-induced ram-pressure stripping \citep{merluzzi2024ram}, suggesting that the processes responsible for diffuse radio emission on Mpc scales may also influence the properties and evolution of individual galaxies.

Among the many tracers used to describe galaxy properties and their evolution (e.g. molecular and ionized gas), neutral atomic hydrogen (\textsc{Hi}) plays a central role. It is the dominant component of the interstellar medium and a key reservoir for star formation. Owing to its typically extended distribution, which on average reaches well beyond the optical disc by a factor of $\sim$2--4 \citep[e.g.][]{wong2002relationship,bigiel2008star,casasola2017radial}, \textsc{Hi} traces the interface between galaxies and their surrounding environment. It is therefore a fundamental probe for studying galaxy evolution and hydrodynamical interactions between the intra-cluster medium and the interstellar medium (e.g. starvation and ram-pressure stripping), which reduce the gas reservoir of galaxies and consequently suppress star formation.

Recent MeerKAT studies have provided new insights into how environment influences galaxy evolution through its impact on \textsc{Hi} content. For example, \citet{sorgho2025meerkat} find that galaxies in evolved Hickson Compact Groups exhibit $\sim$1.5 dex stronger \textsc{Hi} deficiencies than those in less evolved systems, while \citet{kleiner2023meerkat} show that dwarf galaxies in the Fornax cluster rapidly lose their gas upon infall into the cluster potential. On larger scales, the MHONGOOSE survey \citep{de2024mhongoose} reveals extended low-column-density \textsc{Hi} reservoirs around nearby galaxies, whereas the ViCTORIA project identifies clear signatures of ram-pressure stripping in Virgo cluster galaxies \citep{boselli2023victoria}. The MIGHTEE-\textsc{Hi} survey has also established a benchmark \textsc{Hi} mass function for field galaxies out to $z\sim0.08$ \citep{ponomareva2023mightee}.

In the SSC, \citet{gwebushe2026neutral} used MeerKAT observations of the dense environment encompassing Abell\,3558 and SC\,1329--313 to show a marked predominance of transition-zone galaxies. This contrasts with the field-dominated xGASS sample, which is primarily composed of star-forming main-sequence galaxies \citep{janowiecki2020xgass}. Their results suggest that galaxies in the SSC predominantly undergo environmental quenching through starvation or strangulation, rather than rapid gas stripping. 

To broaden our understanding of the accretion processes occurring within the SSC, we investigate the \textsc{Hi} content of galaxies across a wider range of environments, with the aim of assessing the role of environment in galaxy evolution. This work builds upon the study of \citet{gwebushe2026neutral}, which focused on the \textsc{Hi} properties of galaxies in the dense core region between Abell\,3558 and Abell\,3562. Here, we extend the analysis to the filamentary structures that connect the SSC--core to the northern and southern clusters Abell\,3559 and Abell\,3560, respectively. This study combines the new MeerKAT observations of the filament regions with archival MeerKAT data from the MeerKAT Galaxy Cluster Legacy Survey \citep[MGCLS;][]{knowles2022meerkat}.  The \textsc{Hi} data are coupled with the redshift coverage of the SSC provided by the Shapley Supercluster Survey project \citep[ShaSS, ][]{merluzzi2015shapley,haines2018shapley}. 

The paper is organised as follows. In Sect.~\ref{sec: Observations and Data Reduction} we present the \textsc{Hi} observations and data reduction; in Sect.~\ref{sec:Detection_HI} we present the source finding algorithm, the cross-check with the optical catalogue, and the final sample of \textsc{Hi} catalogue; in Sect.~\ref{sec:Results} we present our results; and in Sect.~\ref{sec:Discussion} we discuss them. A summary of the main findings, conclusions, and future prospects is given in Sect.~\ref{sec:Summary_Conclusions}.

Throughout the paper we assume a $\Lambda$CDM cosmology, with H$_0$=70 km/s and $\Omega_{\rm M}$=0.3. At the average redshift of the SSC, $\left<z\right>=0.048$,
1 arcsec=0.947 kpc.
       
\section{ \ion{H} {i} Observations and Data Reduction} 
\label{sec: Observations and Data Reduction}
\begin{figure}
	
	\includegraphics[angle=0,width=\columnwidth]{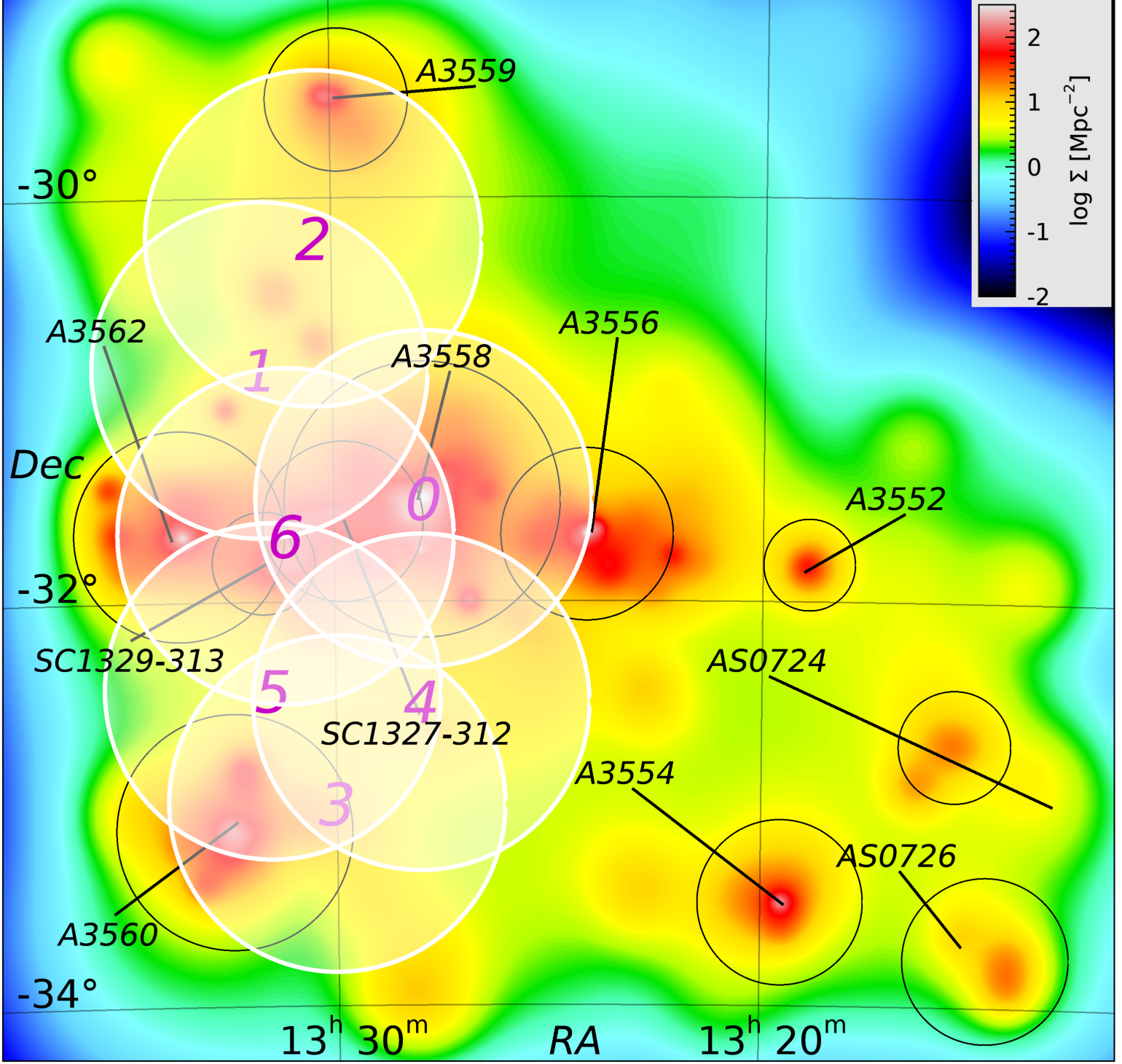}
    \caption{Map of the projected galaxy density of the SSC members identified through spectroscopic redshifts in the Shapley Supercluster Survey region \citep{merluzzi2015shapley}. The 11 clusters within the surveyed area are indicated by black circles with radii corresponding to $\textit{R}_{200}$. The white circles show the seven MeerKAT \textsc{Hi} pointings analysed in this paper. Numbers 0 and 6 mark the centres of the MGCLS pointings, while numbers 1 to 5 mark the new pointings covering the northern and southern filaments connected to the supercluster core.}
    \label{fig:veturi_2022_FoV}
\end{figure}

We combine new and previously published data to investigate the \textsc{Hi} content in the core region of the SSC. The archival data comprise L--band MeerKAT visibilities from the MeerKAT Galaxy Cluster Legacy Survey \citep[MGCLS;][]{knowles2022meerkat}, including observations of the dense cluster Abell\,3558 and the poor cluster SC\,1329$-$313 \citep{venturi2022radio} that were recorded using the 4k channelisation mode. A summary of these observations is provided in Table~\ref{tab:Observations}, with additional details given in \citet{venturi2022radio}. These two pointings are indicated as 0 and 6 in Fig.~\ref{fig:veturi_2022_FoV}. The reduction of the archival MGCLS datasets is described in \citet{knowles2022meerkat}, and the resulting \textsc{Hi} cube properties are summarised in Table~\ref{tab:HI_properties}; further details are provided in \citet{gwebushe2026neutral}.

The new \textsc{Hi} data consist of five MeerKAT pointings (data ID: SCI–20241101–LG–01) targeting the filamentary structure of the SSC. These observations trace the northern and southern filaments connecting the supercluster core to Abell\,3559 in the north and Abell\,3560 in the south. The corresponding pointings are labelled 1--5 in Fig.~\ref{fig:veturi_2022_FoV}, and their observational parameters are summarised in Table~\ref{tab:Observations}. The white circles in Fig.~\ref{fig:veturi_2022_FoV} indicate the field of view (FoV) of the pointings. A smaller radius (i.e. FoV$\sim 50\arcmin$) was used for the \textsc{Hi} analysis to avoid increased noise at the image edges after primary beam correction. The data processing of the five new filament pointings is described in detail in Sect.~\ref{sebsec: HI data processing}.

\begin{table*}
\centering
\caption{Logs of the Observations.}
\label{tab:Observations}
\renewcommand{\arraystretch}{0.9}
\begin{tabular}{lccccclccccc}

\hline \hline
Target ID &
RA &
Dec &
Array &
Project &
Obs. Date & 
$\nu$ & $\Delta\nu _{\rm BW}$& 
$\Delta\nu$ & 
$t_{\rm  int}$ & $N_{\rm ant}$ & $N_{\rm chan}$\\ \\
&
(h:m:s) &
($^{\circ}$:$^{\prime}$:$^{\prime\prime}$) &
&
 &
& [MHz] & [MHz] & [kHz] & [h] &  &\\
\hline \\

Abell\,3558  (P0)   & 13:27:54 & $-$31:29:32 & MeerKAT & MGCLS & 04-Jul-18 & 1284 & 856& 209 & 7 & 61 & 4k\\
SC\,1329$-$313 (P6) & 13:31:08 & $-$31:40:23 & MeerKAT & MGCLS & 06-Jul-18 & 1284 & 856 & 209 &8 & 59 & 4k \\
N film $-$1 (P1)   & 13:31:41 & $-$30:51:00 & MeerKAT & SCI$-$20241101$-$LG-01 & 15-Mar-25 & 1284 & 856 & 26 & 9 & 61 & 32k \\
N film $-$2 (P2)   & 13:30:24 & $-$30:12:00 & MeerKAT & SCI$-$20241101$-$LG-01 & 18-Mar-25 & 1284 & 856 & 26 & 9 & 60 & 32k\\
S film $-$1 (P3)   & 13:30:00 & $-$33:00:00 & MeerKAT & SCI$-$20241101$-$LG-01 & 08-Mar-25 & 1284 & 856 & 26 & 9 & 62 & 32k\\
S film $-$2 (P4)   & 13:28:00 & $-$32:30:00 & MeerKAT & SCI$-$20241101$-$LG-01 & 20-Mar-25 & 1284 & 856 & 26 & 9 & 62 & 32k\\
S film $-$3 (P5)   & 13:31:29 & $-$31:29:32 & MeerKAT & SCI$-$20241101$-$LG-01 & 01-Apr-25 & 1284 & 856 & 26 & 9 & 58 & 32k\\

\hline
\hline
\end{tabular}

\vspace{2mm}
\begin{center}
\textit{Notes}: $\nu$ and $\Delta\nu _{\rm BW}$ refer to the central frequency and total bandwidth respectively, while $\Delta\nu$ refers to channel width.  
\end{center}

\end{table*}

\begin{table*}
    \noindent
	\caption{\textsc{Hi} properties for all data cubes.}
	\label{tab:HI_properties}
	\begin{tabular}{lccccccc} 
		\hline \hline
		Target ID & A\,3558 (P0) & SC\,1329 (P6) & N film$-$1 (P1) & N film$-$2 (P2) & S film$-$1 (P3) & S film$-$2 (P4) &  S film$-$3 (P5)\\
		\hline \\
		RA (J2000) &13$^\textup{h}$ 27$^\textup{m}$ 54$^\textup{s}$ &13$^\textup{h}$ 31$^\textup{m}$ 08$^\textup{s}$ & 13$^\textup{h}$ 31$^\textup{m}$ 42$^\textup{s}$ &13$^\textup{h}$ 30$^\textup{m}$ 24$^\textup{s}$& 13$^\textup{h}$ 30$^\textup{m}$ 00$^\textup{s}$ &13$^\textup{h}$ 28$^\textup{m}$ 00$^\textup{s}$ &13$^\textup{h}$ 31$^\textup{m}$ 29$^\textup{s}$\\
            Dec (J2000) & $-$31$\degr$ 29$\arcmin$ 32$\arcsec$ & $-$31$\degr$ 40$\arcmin$ 23$\arcsec$ & $-$31$\degr$ 51$\arcmin$ 00$\arcsec$ & $-$31$\degr$ 12$\arcmin$ 00$\arcsec$ & $-$33$\degr$ 00$\arcmin$ 00$\arcsec$ & $-$32$\degr$ 30$\arcmin$ 00$\arcsec$ & $-$31$\degr$ 29$\arcmin$ 32$\arcsec$\\
		$V_{h}$ [km/s] &$14500\pm39$  &$13416\pm49$ & \dots & \dots & \dots & \dots & \dots  
        \\
		  $\sigma_{v}$ [km/s] &$1007\pm25$ &$373 \pm28$ & \dots & \dots & \dots & \dots & \dots \\
            $z$ &0.048 &0.045 &0.048 &0.048 &0.048 &0.048 &0.048\\
            $\textit{R}_{200}$ [Mpc] &2.32  &0.862 & \dots & \dots & \dots & \dots & \dots \\
            $\textit{M}_{200}\ [\textup{h}^{-1} \textup{M}_{\odot}]$ &$4.47^{+ 2.78}_{-2.38}\times 10^{14}$ &$1.79^{+ 2.93} \times 10^{13}$ & \dots & \dots & \dots & \dots & \dots\\            
            RMS [mJy/beam] & $\sim 0.22$ &$\sim 0.22$ & $\sim 0.21$ & $\sim 0.21$ & $\sim 0.21$ & $\sim 0.21$ & $\sim 0.21$ \\ 
            BMAJ [arcsec] &33.05   &34.75 & 33.40 & 33.72 & 33.70 & 33.75 & 33.49\\
            BMIN [arcsec] &30.41   &30.96 & 30.03 & 30.18 & 30.35 & 30.38 & 32.34 \\
            BPA [deg]     & $-$29.51  &$-$53.48 & $-$61.68 & $-$51.74   &  $-$57.01   &  $-$49.84  & $-$63.36  \\
		\hline
        \hline
	\end{tabular}
 \\
 \textit{Notes}:  The central velocities ($V_{h}$), velocity dispersions ($\sigma_{v}$), and $\textit{R}_{200}$ radii values were obtained from \citet{haines2018shapley}, while the $\textit{M}_{200}$ masses values were derived from the work of \citet{higuchi2020shapley}. 
\end{table*}

\subsection{\textsc{Hi} data processing} 
\label{sebsec: HI data processing}
\subsubsection{\rm \textbf{Observations}}

The SSC filaments were observed with the MeerKAT radio telescope in L--band spectral--line mode as part of project SCI--20241101--LG--01. The observations targeted five pointings sampling the northern and southern filamentary structures surrounding the supercluster core. The locations and observational parameters of the individual pointings are summarised in Table~\ref{tab:Observations}.

The observations were conducted between 8 March and 1 April 2025 using between 58 and 62 MeerKAT antennas. Each pointing was observed for a total of 9 h, including $\sim 2\ \text{h}$ for overhead calibration, resulting in a total observing time of 45 h. The observations were centred at 1284 MHz with a total bandwidth of 856 MHz. Data were recorded using the 32k channelisation mode, providing a velocity resolution of 5.76 km/s ($\sim$ 26 kHz) at the redshifted \textsc{Hi} frequency of the SSC, and only parallel-hand correlations were recorded with an integration time of 8 s.
The observations were scheduled during night-time to minimise the impact of radio frequency interference (RFI), thereby improving the sensitivity to low-surface-brightness and diffuse \textsc{Hi} emission. 

\subsubsection{\rm \textbf  {\textsc{Hi} Data Reduction}}

The MeerKAT visibility data were transferred from the SARAO archive to the Ilifu%
\footnote{\textcolor{blue}{\url{https://www.ilifu.ac.za/}}} cloud computing facility for processing. During the extraction, standard archive flags associated with telescope status (\texttt{cam}), data loss, and radio-frequency interference (RFI) identified during data ingestion were applied. In addition, the first second of each scan was flagged to remove potential start-up transients. Level--1 calibration solutions (i.e. transfer calibration, also known as 1GC) were applied during extraction, and all four polarisation products (HH, VV, HV, and VH) were retained. The data were recorded with an integration time of 8 s.

To optimise storage requirements and processing time, only channels 18,528--19,767 were extracted from the archive and averaged by a factor of two in frequency. The channel averaging results in 620 channels. The selected channel range corresponds to the redshifted \textsc{Hi} emission over the SSC redshift interval (0.03 $< z <$ 0.06), equivalent to an observed frequency range of approximately 1340--1379 MHz and a recessional velocity range of $\sim$ 9000--18 000\ \rm km/s. This selection encompasses the full velocity range of the supercluster while excluding most foreground and background regions outside the volume of interest. Edge channels affected by bandpass roll-off were also discarded to ensure a uniform spectral response across the band. The channel averaging significantly reduced the data volume and computational requirements of the subsequent calibration, imaging, and source-finding procedures while preserving the spectral coverage required for the \textsc{Hi} study of galaxies in the SSC.

We reduce the data using the Containerised Automated Radio Astronomical Calibration pipeline%
\footnote{\textcolor{blue}{\url{https://caracal.readthedocs.io}}} \citep[\texttt{CARACal}; ][]{jozsa2020caracal}. \texttt{CARACal} allows users to process radio interferometric data using a range of software packages within a single pipeline through Stimela%
\footnote{\textcolor{blue}{\url{https://github.com/SpheMakh/Stimela}}} \citep[][]{Makhathini2018, smirnov2025africanus}, a platform-independent radio interferometry scripting framework based on Python and container technology \citep[in our case, Singularity;][]{kurtzer2017singularity}. The framework supports the integration of various open-source radio interferometry software packages within the same processing workflow.

The Level--1 calibration solutions were applied during the extraction from the SARAO archive, and the calibrator and target visibilities were already flagged and cross-calibrated. In addition to the archive flags, further RFI flagging was applied to the target visibilities using AOFlagger \citep{offringa2010post,offringa2012morphological}, through the \texttt{flag\_mst\_errors} step in \texttt{CARACal}, with the \texttt{postmst.rfis} strategy. The first step in our data processing was therefore to obtain a continuum model and perform self-calibration prior to continuum subtraction and \textsc{Hi} imaging.

To ensure accurate continuum subtraction, we created a model of the continuum emission by imaging and self-calibrating the target field iteratively. Imaging was performed with \texttt{WSClean} \citep[][]{offringa2014wsclean,offringa2017optimized} using Briggs weighting with a \textit{robust} parameter of $\rm r=-0.5$. Cleaning was performed down to 0.5$\sigma$ within masks generated with \texttt{SoFiA} \citep[][]{serra2015sofia,westmeier2021sofia}. We self-calibrated the gain phase with \texttt{CubiCal} \citep{kenyon2018cubical} using solution intervals of approximately 120 s, and two rounds of self-calibration were performed. The resulting continuum model was subtracted from the calibrated visibilities. Residual continuum emission was further reduced by fitting a third-order polynomial to the visibilities on the \textsc{Hi} line free channels.

Continuum subtraction was performed in the uv-plane using a third-order polynomial fit (\texttt{uvlin}). The continuum-subtracted uv-data were subsequently Doppler-corrected and Fourier transformed into the final \textsc{Hi} data cube using \texttt{WSClean}. Deconvolution was performed using automatic masking, with an initial masking threshold of $10\sigma$ and a final cleaning threshold of $0.5\sigma$. The maximum number of CLEAN iterations was set to $10^6$, with a major-cycle loop gain of 0.9. 

Imaging was performed using Briggs weighting with a \textit{robust} parameter of $\rm r=0.5$ and 20$\arcsec$ \texttt{uv}-tapering to optimise the surface brightness sensitivity (this produces cubes with a synthesized beam (angular resolution) of $\sim 30\arcsec$, similar to the cluster cubes). The primary beam response was corrected in the final cube using the MeerKAT primary beam model. 
The resulting \textsc{Hi} cubes consist of 152 spectral channels (after binning/averaging our cubes from $\sim 32$k mode to $\sim 4$k mode, in order to achieve similar channel width, $\sim 209 \ \text{kHz or 44 km/s}$, between the cluster centre cubes \citep{gwebushe2026neutral} and the new filament cubes) covering a frequency range of 1.3401--1.3718 GHz, corresponding to a velocity range of 10,251--16,956 km/s. The image properties of each cube, including restoring beam parameters and rms noise levels, are summarised in Table~\ref{tab:HI_properties}. 

\begin{table*}
    \centering
    \caption{Calculated brightness temperature noise and $3\sigma$ \textsc{Hi} column density sensitivities.}
    \label{tab:column_density_sensitivity}
    \begin{tabular}{lccccc}
        \hline \hline
        Target ID & Beam Dimensions & RMS Noise & Channel Width & $\sigma_{T_{\textup{B}}}$ Noise & $3\sigma$ $N_{\textsc{Hi}}$ Sensitivity \\
         & ($\textup{BMAJ} \times \textup{BMIN}$) & [mJy/beam] & [km/s] & [K] & [$\textup{atoms~cm}^{-2}$] \\
        \hline \\
        Abell\,3558 (P0)      & $33.03\arcsec \times 30.42\arcsec$ & $0.22$ & $44$ & $0.1464$ & $3.52 \times 10^{19}$ \\
        SC\,1329-313 (P6)    & $34.75\arcsec \times 30.96\arcsec$ & $0.22$ & $44$ & $0.1357$ & $3.27 \times 10^{19}$ \\
        N film$-$1 (P1) & $33.40\arcsec \times 30.03\arcsec$& $0.21$ & $44$ & $0.1268$ & $3.05 \times 10^{19}$ \\
        N film$-$2 (P2) & $33.72\arcsec \times 30.18\arcsec$ & $0.21$ & $44$ & $0.1250$ & $3.01 \times 10^{19}$ \\
        S film$-$1 (P3) & $33.70\arcsec \times 30.35\arcsec$ & $0.21$ & $44$ & $0.1244$ & $2.99 \times 10^{19}$ \\
        S film$-$2 (P4) & $33.75\arcsec \times 30.38\arcsec$ & $0.21$ & $44$ & $0.1241$ & $2.99 \times 10^{19}$ \\
        S film$-$3 (P5) & $33.49\arcsec \times 32.34\arcsec$ & $0.21$ & $44$ & $0.1174$ & $2.83 \times 10^{19}$ \\
        \hline \hline
    \end{tabular}
\end{table*}


Using the \textsc{Hi} cube properties listed in Table~\ref{tab:HI_properties}, we estimate the 3$\sigma$ column density sensitivity, $N_{\rm \textsc{Hi}}$, for each cube. The column density sensitivity is calculated following \citet{meyer2017tracing}:

\begin{equation}\label{eq:N_HI_equ}
   N_{\textsc{Hi}} = 1.823\times10^{18} \int T_{B}\ \textup{d}v \approx 1.823\times10^{18} \times(3 \cdot \sigma_{T_{B}} )\times \Delta v,
\end{equation} 

\noindent
where $N_{\textsc{Hi}}$ is in units of $\rm atoms\ cm^{-2}$, $\Delta v$ is the per-channel width (i.e. $44.11\,\rm km/s$ for both cluster and filament cubes), and $\sigma_{T_{B}}$ is the per-channel brightness temperature rms noise. The latter is derived from the beam dimensions, BMAJ and BMIN, and the flux-density rms noise, $S_v=\sigma$, using the Rayleigh--Jeans approximation \citep{meyer2017tracing}:

\begin{equation}\label{eq:T_b_equ}
   \sigma_{T_{B}} = \frac{1.222\times10^6}{\text{BMAJ} \cdot \text{BMIN} \cdot \nu^2} \cdot S_{v},
\end{equation} 

\noindent
where $\nu$ is the observed frequency in GHz. The resulting column density sensitivities are summarised in Table~\ref{tab:column_density_sensitivity}. The cluster and filament \textsc{Hi} cubes were imaged with comparable angular resolutions and a common spectral resolution. The resulting rms noise levels are $\sim0.22$ mJy/beam for the cluster cubes and $\sim0.21$ mJy/beam for the filament cubes. These similar angular, spectral, and noise properties provide comparable \textsc{Hi} sensitivities for the two environments. 


\section{Detection of \ion{H}{i} and sample selection} 
\label{sec:Detection_HI}

We make use of the \textsc{Hi} data products presented in \citet{gwebushe2026neutral} for the MGCLS pointings centred on A\,3558 and SC\,1329--313 in the SSC-core. Details of i) the \textsc{Hi} source finding; ii) the full and complete \textsc{Hi} sample; iii) the cross-correlation between the \textsc{Hi} detections and the ShaSS spectroscopic catalogue \citep[][]{haines2018shapley} and  iv) the morphological classification of the \textsc{Hi}-detected SSC-core members are presented in \citet{gwebushe2026neutral}. In addition to that work, we estimate \textsc{Hi} upper limits for non-detected galaxies in both the SSC clusters and the filament regions. In Subsection~\ref{sebsec: Filaments pointings}, we follow a similar procedure to derive the corresponding data products for the newly observed SSC filament pointings.

\subsection{Filament pointings around the SSC-core} 
\label{sebsec: Filaments pointings}

\subsubsection{\rm \textbf{Source finding}}
\label{sec:Source_finding}

We utilise the Source Finding Application 2 pipeline \citep[SoFiA 2,][]{serra2015sofia, westmeier2021sofia} to search for \textsc{Hi} emission sources within the redshift interval $z \sim 0.0354-0.0599$ (corresponding to $cz \sim 10251-16956$ km/s) using the SoFiA smooth$+$clip (S$+$C) source finder. Prior to source detection, the data cubes are preconditioned by (1) estimating and normalising the local noise over a running window of $25 \times 25$ spatial pixels and 15 spectral channels using the median absolute deviation (MAD) statistic, and (2) identifying and flagging outlier voxels with absolute values greater than $5$ times the local noise level. This step reduces the impact of residual artefacts that could affect the reliability of the source-finding process.

The S$+$C algorithm is applied using Gaussian spatial kernels with sizes of 0, 3, 6, and 9 pixels, combined with spectral boxcar kernels of 0, 3, and 7 channels. We adopt a detection threshold of 3.8$\sigma$, where voxels above this threshold are assigned a replacement value of 2$\sigma$. This allows detections identified across the different spatial and spectral smoothing scales to be combined into a single final detection mask.

Detections are linked using a spatial linking radius of 2 pixels and a spectral linking radius of 3 channels, with sources required to have a minimum extent of 3 spatial pixels and 3 spectral channels to be considered reliable. We then apply the SoFiA reliability filter to remove detections with reliability values below 0.85. Within the SoFiA framework, the reliability parameter represents the probability that a detection corresponds to a genuine astronomical source rather than a noise fluctuation. This is estimated statistically by comparing the distributions of positive and negative detections in the parameter space defined by the peak flux, integrated flux, and mean flux of the candidate sources. Therefore, applying a reliability threshold of 0.85 retains detections with an estimated probability of $\geq 85$ per cent of being real. Additionally, we require a minimum signal-to-noise ratio (SNR) of 3.2. The remaining detections are then parameterised assuming a restoring beam size of approximately 30$\arcsec$ for all integrated flux measurements.

After removing artefacts, such as those associated with bandpass ripple, the final catalogue for the filament pointings contains 232 \textsc{Hi} detections with an integrated SNR $\geq$ 3.2. We adopt a $3\sigma$ detection threshold for the filament pointings. The per-channel spacing of the data cubes is $\sim44$ km/s. To minimise the inclusion of spurious detections, such as RFI or noise fluctuations, we require all detected sources to extend over at least three consecutive spectral channels (i.e. $\gtrsim3\times44$ km/s). Under these criteria, the resulting lowest-detected \textsc{Hi} mass is 
$M_{\textsc{Hi}} \sim 3.11 \times 10^{8}\,\mathrm{M}_{\odot}$ located at $z_{\textsc{Hi}} \sim 0.038$. This corresponds to a column density sensitivity of $N_{\textsc{Hi}} \sim 4.59 \times 10^{19}\,\mathrm{atoms\ cm}^{-2}$. 
\subsubsection{\rm \textbf{Cross-Matching}}
\label{sec:Cross_correlation}

Our \textsc{Hi} catalogue was cross-correlated with the ShaSS spectroscopic catalogue, which is based on observations obtained with ESO-VST, AAOmega and literature data. The \textsc{Hi} catalogue contains 232 objects within the area covered by the MeerKAT observations of the SSC filaments. For each \textsc{Hi} detection, we searched for a corresponding optical counterpart in the Visible (hereafter VIS) catalogue of the ShaSS project \citep{merluzzi2015shapley}. The ShaSS spectroscopic catalogue is 95 per cent complete to a magnitude limit of $\textit{i}<18$, corresponding to a stellar mass of approximately $\log_{10}(M_{\star}/\textup{M}_{\odot}) \sim 9.3$, but extends down to $\log_{10}(M_{\star}/\textup{M}_{\odot}) = 8$ \citep{haines2018shapley}.

Of the 232 \textsc{Hi} detected sources, 208 are associated to galaxies in the VIS with spectroscopic redshift available and in the range of the SSC members as derived by \citet{haines2018shapley}. All overlays of the 208 sources are re-inspected, detecting wrong associations, interactions, objects affected by the halos of nearby bright stars, and objects with superimposed stars. We also found double detections and selected the most likely one. This results in a final sample of 142 \textsc{Hi} detections. All \textsc{Hi} detections in the filament pointings are associated with optically identified galaxies in the available catalogue; therefore, no isolated \textsc{Hi} clouds without stellar counterparts are found. 

In addition to the 142 \textsc{Hi}-detected galaxies in the new filament pointings, \citet{gwebushe2026neutral} identified a further 15 \textsc{Hi}-detected galaxies (i.e. in MGCLS pointings) belonging to the SSC that lie outside the $\textit{R}_{200}$ regions of all clusters, based on the dynamical analysis of \citet{haines2018shapley}. These galaxies trace the filament connecting the clusters in the SSC-core. Combining the two samples yields a final filament population of 157 \textsc{Hi}-detected galaxies. As described in \citet{gwebushe2026neutral}, the cluster centre contains 106 \textsc{Hi}-detected galaxies. In addition, 21 \textsc{Hi} sources belonging to cluster centre, based on the dynamical analysis of \citet{haines2018shapley}, were determined in the filament pointings, bringing the total number of \textsc{Hi}-detected galaxies in the cluster centre to 127. 

To summarise, we determined 157 \textsc{Hi}-detected galaxies in the filament and 127 \textsc{Hi}-detected galaxies in the cluster centre.

\subsubsection{\rm \textbf{Stellar Masses and Star-Formation Rates Determination}}

Stellar masses and specific star-formation rates (sSFRs) were derived through SED fitting, by comparing the available ShaSS $ugriK$ photometry (in the AB system) with synthetic photometry for a library of composite stellar population models. This procedure has been applied both to galaxies in all environments including field/filaments and clusters.

Synthetic observed-frame photometry was generated using an updated version of the E-MILES simple stellar population (SSP) models \citep[][\textcolor{blue}{Vazdekis et al. 2026, in prep}]{vazdekis2016uv}. The E-MILES models cover the wavelength range $0.168-5\ \mu\rm m$, combining several empirical stellar libraries: NGSL \citep{gregg2006hst,koekemoer20062005}, MILES \citep{sanchez2006medium}, Indo-US \citep{valdes2004indo}, CaT \citep{cenarro2001empirical} and IRTF \citep{cushing2005infrared,rayner2009infrared} \citep[see also][]{vazdekis2012miuscat,rock2016miles}. In the updated version, the NIR spectral range is based on both the original IRTF library and its extension \citep{villaume2017extended}, improving the coverage of the subsolar and supersolar metallicity regimes \citep[see][for details]{la2024radial}. For the present analysis, we adopted SSPs based on BaSTI isochrones \citep{pietrinferni2004large,vazdekis2015evolutionary}, assuming a bimodal IMF with slope $\Gamma _{b} = 1.30$, which closely resembles the Kroupa IMF.

Each composite stellar population model consists of two components: (i) an old population following an exponentially declining star-formation history (SFH), truncated at a lookback time of 50 Myr, and (ii) a single young SSP of solar metallicity. For the old component, we considered 30 metallicities, $-1.19\leq [\rm M/H] \leq + 0.26$, spaced by 0.05 dex, 19 maximum lookback ages between 4.5 and 13.5 Gyr (in steps of 0.5 Gyr), and 42 exponential timescales spanning approximately $0.1-10$ Gyr. This yields
$30 \times 19 \times 42 = 23 940$ old-component templates. The young component was represented by 31 SSPs with ages between 50 Myr and 3 Gyr. We found that extending the SSP grid towards younger ages or lower metallicities did not improve the agreement with the observed photometry.

Synthetic photometry was computed by integrating the redshifted model SEDs through the SDSS $u$, $g$, and $r$, $i$, and VISTA $K_{s}$ transmission curves. The grid includes 18 redshifts between $z =\rm 0.036\ and\ 0.070$ in steps of 0.002, and 11 reddening values, $ E(B - V) = 0-0.50$, in steps of 0.05 mag. All SSPs were normalized to one solar mass of initially formed stars. The corresponding fractions of surviving stellar mass, $\mu$, were used to derive stellar masses and sSFRs from the best-fitting composite models.

For each galaxy, aperture magnitudes measured within circular diameters of 1.5, 3, 4, 8, 16, 30, 45, and 90 arcsec were interpolated to a diameter $D= 2r_{\rm Kron}$, where $r_{\rm Kron}$ is the circularized Kron radius measured in the $r$ band. To mitigate the effects of unreliable photometry or contamination in large apertures, we adopted an adaptive aperture-selection procedure. For galaxies with $\chi^2> 15$, the fit was repeated at successively smaller radii of 0.8, 0.6, and 0.5$r_{\rm Kron}$, imposing a minimum aperture diameter of 3 arcsec to limit seeing-related effects. The first aperture yielding $\chi^2\leq 15$ was adopted, or the smallest allowed aperture if this criterion was not met. However, if the resulting $\rm sSFR_{1.0}$ differed by no more than 0.15 dex from the value obtained with valid full-Kron photometry, we retained the full-Kron solution regardless of its $\chi^2$. This procedure limits the influence of corrupted or contaminated large-aperture measurements while retaining the largest possible aperture whenever the inferred star formation activity remains consistent. Approximately 10 per cent of the galaxies were ultimately fitted using apertures smaller than the full Kron aperture. 

To place the aperture photometry on an approximately total-flux scale, we applied a common magnitude offset to all available bands, computed as the mean difference between the Kron and interpolated aperture magnitudes in the $r$, $i$, and $K$ bands.

At each galaxy redshift, the synthetic fluxes were linearly interpolated between adjacent redshift-grid nodes. For each reddening value and combination of old and young templates, we fitted the observed five-band fluxes with a non-negative linear combination of the two components. Their amplitudes were determined analytically by weighted least squares, and the minimum $\chi^2$ was identified through an exhaustive search of the model grid.

Photometric uncertainties were combined in quadrature with minimum magnitude errors of 0.028 mag in the $u$ band and 0.020 mag in the other bands. The fitted amplitudes determine the relative contributions of the old and young populations, while stellar masses were derived from the overall five-band best-fitting normalization, accounting for the surviving stellar-mass fraction of the composite population.

The sSFR averaged over a lookback interval $\Delta t$ was calculated as

\begin{equation}\label{eq: sSFR_equ}
   \rm sSFR_{\Delta \it t} = \frac{\it f_{\rm formed}\rm(<\Delta \it t)}{\mu \Delta \it t} ,
\end{equation} 
where $f_{\rm formed}\rm(<\Delta \it t)$ is the fraction of the initially formed stellar mass produced within the last $\Delta \it t$, and $\mu$ is the fraction of surviving stellar mass. Here, $\Delta \it t$ is expressed in years. We considered $\Delta \it t = \rm 0.5,\ 1.0,\ and\ 1.5 \ Gyr$, and refer to the corresponding quantities as $\rm sSFR _{0.5}, \ sSFR _{1.0} \ and\ sSFR _{1.5}$, respectively. In the present analysis we use $\rm sSFR _{1.5}$ because it yelds a distribution of field/filament galaxies with normal \textsc{Hi} content that most closely resembles that of xGASS \citep{janowiecki2020xgass}.

Parameter uncertainties were estimated from 200 Monte Carlo realizations per galaxy. For each realization, the best-fitting model photometry was perturbed with Gaussian noise, accounting for the observational magnitude errors and the adopted error floors. To allow for additional photometric scatter in galaxies with poor fits, the total effective uncertainties were multiplied by

\begin{equation}\label{eq: photometric_scatter}
   s_{i} = \left[ \rm max \left(1, \frac{\chi^2_{\it i}}{\widetilde{\chi^2}} \right)\right],
\end{equation} 

where $\chi^2_{\it i}$ is the best-fitting value for galaxy $i$, and $\widetilde{\chi^2}$ is the median over the sample. Each simulated data set was refitted using the same model grid, and the resulting parameter distributions were characterized by their 16th and 84th percentiles. The 16th–84th percentile range was adopted as the uncertainty interval.

The \textsc{Hi}-detected galaxies in the filament cover a stellar mass range of $8.10<\rm log_{10}\left(\it M_{\star}/\rm M_{\odot}\right)< 11.46$, while the cluster centre cover  $8.39<\rm log_{10}\left(\it M_{\star}/\rm M_{\odot}\right)< 11.44$. 

\subsubsection{\rm \textbf{Spatial Distribution, Phase-Space, and Environmental Classification}}
\label{sec:Statistical_properties}

\begin{figure*}
    \centering
    \includegraphics[width=\textwidth]{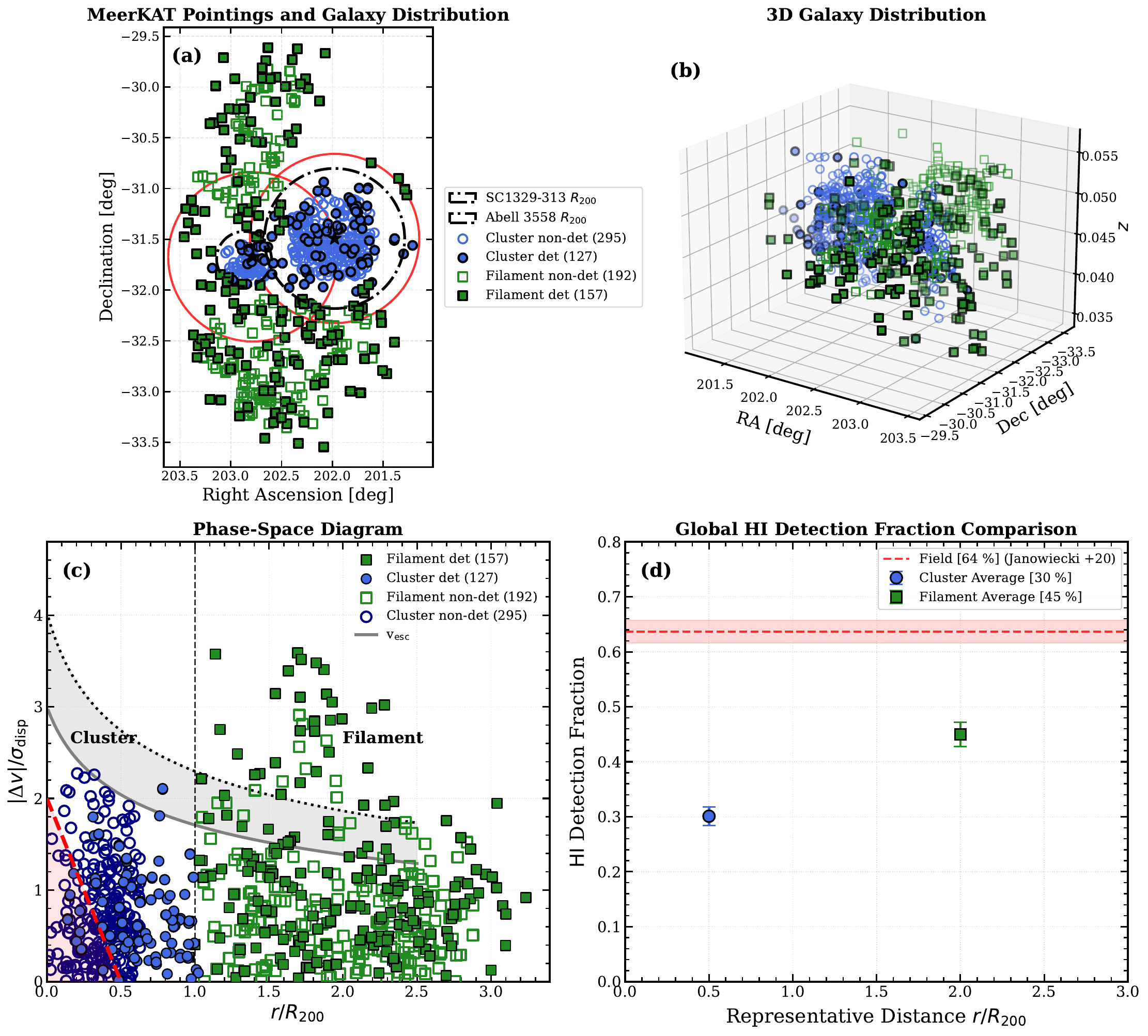}
    \caption{Spatial, kinematic, and statistical distributions of the galaxy sample across different environments. 
    \textit{Top-left (a):} Projected spatial distribution in Right Ascension and Declination. Large red circles denote the FoV$\sim 50 \arcmin$ of the pointings centered at A\,3558 and SC\,1329. Dash-dotted black lines represent the cluster virial boundaries ($R_{200}$) for SC\,1329 and A\,3558. \textsc{Hi} detections and non-detections are represented by filled and open markers, respectively, split into the cluster (blue circles) and filament (green squares) populations. 
    \textit{Top-right (b):} Three-dimensional spatial distribution showing the structure of the cosmic web as a function of Right Ascension, Declination, and redshift ($z$). 
    \textit{Bottom-left (c):} Projected phase-space diagram showing normalized relative line-of-sight velocity ($|\Delta v|/\sigma_{\rm disp}$) versus normalized cluster-centric radius ($r/R_{200}$). The solid and dotted black lines trace the analytical escape velocity, $v_{\rm esc}$, and 3$\sigma_{\rm esc}$ profiles, where $\sigma_{\rm esc}$ is the uncertainty in the escape velocity. The dashed red line bounding the light red shaded area marks the virialized region following \citet{rhee2017phase}. The vertical dashed line at $r/R_{200} = 1$ separates the high-density cluster centre from the infalling filamentary environment. 
    \textit{Bottom-right (d):} Global \textsc{Hi} detection fractions computed for the cluster (blue circle) and filament (green square) populations, plotted with binomial uncertainty bars. The horizontal red dashed lines correspond to the field reference detection fraction from  \citet{janowiecki2020xgass}, with the associated error range highlighted by the shaded region.}
    \label{fig:sample_overview_2x2}
\end{figure*}

The spatial, phase-space, and statistical properties of our galaxy sample are structured into a multi-panel distribution in Fig.~\ref{fig:sample_overview_2x2}. 

The top row presents the distribution of the sample. Fig.~\ref{fig:sample_overview_2x2}a displays the projected sky distribution of the galaxies relative to the two MGCLS pointings (i.e. large red circles represent the FoV) and the cluster virial boundaries ($ R_{200}$) are shown by dash-dotted black circles. In this panel, \textsc{Hi} detections and non-detections of SSC members are represented by filled and open markers, respectively, and are color-coded by their environment: blue circles for the cluster environment and green squares for the filament environment. This environmental division is extended into three dimensions in Fig.~\ref{fig:sample_overview_2x2}b, which shows the distribution of the sample across Right Ascension, Declination, and redshift ($z$) space.

The bottom row displays the phase-space diagram and the \textsc{Hi} detection fraction as a function of the normalized cluster-centric radius ($r/R_{200}$), centered on A\,3558. Fig.~\ref{fig:sample_overview_2x2}c presents the projected phase-space diagram mapping the relative line-of-sight velocity ($|\Delta v|/\sigma_{\rm disp}$) against $r/R_{200}$. Cluster member galaxies are defined by projected radial distances of $r \leq R_{200}$ and relative line-of-sight velocities positioning them below the cluster escape velocity envelope (indicated by the dotted black curve). The dashed red line, which bounds the light red shaded region, delineates the virialised region \citep[][]{mahajan2011velocity,jaffe2015budhies} containing 9 ($\sim$7 per cent) \textsc{Hi}-detected galaxies and 80 ($\sim$27 per cent) \textsc{Hi} non-detected galaxies. We calculate the cluster escape velocity and its associated $3\sigma_{\rm esc}$ uncertainty profile ($\sigma_{\rm esc}$) following Equations~(1)--(4) from \citet{rhee2017phase}, using the global parameters $R_{200}$, $ M_{200}$, and $\sigma_{\rm disp}$ summarized in Table~\ref{tab:HI_properties}. Infalling cluster populations are defined within the regime $1 < r/R_{200} \leq 2.5$ bound below the escape velocity curve, while galaxies outside these bounds are classified as the field population \citep{rhee2017phase,reynolds2022wallaby}. 

However, because the SSC is a dense and dynamically complex supercluster, separating galaxies into "infall" and "field" populations is ambiguous. For example, a galaxy located on the outskirts of A\,3558 may appear to be a field galaxy with respect to that cluster, while at the same time belonging to the infall region relative to the adjacent A\,3559 or A\,3560 structures. To avoid introducing classification biases into our analysis, we classify all non-cluster centre galaxies ($r > R_{200} \ \text{and/or}\ |\Delta v|/\sigma_{\rm disp} \geq 3\sigma_{\rm esc}$, but having redshift consistent with the SSC spectroscopic range) as belonging to a single "filament" population, rather than distinguishing between individual infall or field populations \citep[see also the SSC dynamical analysis by][]{haines2018shapley}.

Finally, Fig.~\ref{fig:sample_overview_2x2}d shows the global \textsc{Hi} detection fractions for the two SSC environments, positioned at representative radii of $r/R_{200} = 0.5$ for the cluster average ($30\ \text{per cent}$) and $r/R_{200} = 2$ for the filament average ($45\ \text{per cent}$). For reference, the \textsc{Hi} detection fraction of $64\ \text{per cent}$ from the xGASS survey is shown as a horizontal line \citep{janowiecki2020xgass}. Given the different sample selection and sensitivities of the surveys, the xGASS value is used as a reference rather than as a directly matched comparison.


\subsection{ Morphological classification of the sample}
\label{sec:Morphological classification}

Optical morphological classifications (early-type, late-type, and irregular) for all ShaSS members were performed following the visual inspection scheme described in \citet{gwebushe2026neutral} \citep[based on][]{2013pss6.book....1B, 2017MNRAS.471.4027B}. Briefly, $r$- and $g$-band imaging was used to evaluate basic galaxy types as well as structural features such as rings, bars, and spiral arms.

Late-type galaxies (LTGs) dominate the \textsc{Hi}-detected samples in both the cluster and filament environments. Table~\ref{tab:morphological} summarises the morphological composition of the samples. The filament detections contain a higher fraction of LTGs (72 per cent) compared to the cluster detections (65 per cent), following the morphology density relation of \citet{dressler1980galaxy}. In contrast, the cluster population contains a larger fraction of early-type galaxies (ETGs), with 67 per cent of ETGs compared to 41 per cent in the filament population when considering both \textsc{Hi} detections and non-detections.

Since \textsc{Hi} observations tend to privilege the detection of \textsc{Hi}-rich galaxies, which are usually LTG, and may fail to detect the \textsc{Hi} poorer ETG, \textsc{Hi}-detected galaxies alone do not provide a complete representation of the underlying galaxy population required to investigate the impact of environment on \textsc{Hi} content and star formation.

\begin{table*}
    \centering
    \caption{Morphological breakdown of the \textsc{Hi} detected and non-detected galaxy populations within the cluster and filament environments. The percentages indicate the fraction of galaxies belonging to each morphological type within that specific environmental column, with the corresponding raw source counts provided in brackets.}
    \label{tab:morphological}
    \small
    \begin{tabular}{l c c c c c}
        \hline \hline
        & \multicolumn{2}{c}\textbf{}{\textbf{SSC--cluster Environment}} & & \multicolumn{2}{c}{\textbf{SSC--filament Environment}}\\
        \hline\\
        \textbf{Morphological Type} & $N_{\rm det}$ & $N_{\rm non-det}$ & & $N_{\rm det}$ & $N_{\rm non-det}$ \\
        
        Late-type galaxies (LTGs)   & 65 per cent [83]  & 11 per cent [32] & & 72 per cent [113]  & 26 per cent [50] \\
        Early-type galaxies (ETGs)  & 18 per cent [23]  &  88 per cent [260]  & & 2 per cent [3]    & 73 per cent [140] \\
        Irregular galaxies (Irr)    & 17 per cent [21]  & 1 per cent [3]    & & 26 per cent [41]  & 1 per cent [2]   \\
        \hline
        \textbf{Total Galaxies}     & \textbf{[127]} & \textbf{[295]} & & \textbf{[157]} & \textbf{[192]} \\
        \hline
        \hline
    \end{tabular}
\end{table*}

\subsection{\textsc{Hi} data products}
\label{sec:HI_data_products}

\subsubsection{\rm \textbf{\textsc{Hi} Mass Determination}}
\label{subsec:hi_Mass}

The total \textsc{Hi} mass, $M_{\textsc{Hi}}$, of the detected galaxies was derived from the total \textsc{Hi} flux, assuming optically thin emission and no self-absorption, using the following formula:

\begin{equation}\label{eq:HI_mass_equ}
   M_{\textsc{Hi}} = 2.356\times10^{5} D_{\textsc{L}}^2 (1+z)^{-1} S_{\textsc{Hi}},
\end{equation} 

\noindent
where $M_{\textsc{Hi}}$ is in units of $\textup{M}_{\odot}$, $D_{\textsc{L}}$ is the luminosity distance of the galaxy in Mpc, $z$ is the redshift of the galaxy, and $S_{\textsc{Hi}}$ is the \textsc{Hi} integrated flux in Jy km\,s$^{-1}$  \citep[see][for details]{meyer2017tracing}. Uncertainties in $M_{\textsc{Hi}}$ have been computed from the uncertainty on the \textsc{Hi} flux.

For galaxies within our sample that lacked a reliable \textsc{Hi} line detection within the MeerKAT FoV, we computed upper limits on their integrated \textsc{Hi} fluxes and corresponding \textsc{Hi} masses. We restricted this analysis to targets located within a conservative FoV radius from the primary pointing centers.

To account for spatial variations in sensitivity, the local root-mean-square (rms) noise ($\sigma_{\rm local}$) was calculated for each galaxy by correcting the per-channel cube sensitivity ($\sigma_0 = 0.22\,\text{mJy}\,\text{beam}^{-1}$ and $\sigma_0 = 0.21\,\text{mJy}\,\text{beam}^{-1}$ for cluster and filament environments, respectively) for primary beam (PB) attenuation:
\begin{equation}
    \sigma_{\rm local} = \frac{\sigma_0}{\mathcal{A}(\theta)},
\end{equation}
where $\mathcal{A}(\theta) = \exp\left[-4 \ln(2) \left(\frac{\theta}{\text{FWHM}_{\rm PB}}\right)^2\right]$ is the primary beam response at an angular offset $\theta$ from the pointing center, assuming a Gaussian profile with a Full Width at Half Maximum ($\text{FWHM}_{\rm PB}$) of $1.2^\circ$.

The integrated flux density upper limit is dependent on the expected spectral line width ($W$) of the source. Following standard practice for unresolved or faint sources, we assume a uniform tophat line profile where the integrated flux density limit scales with the square root of the velocity width \citep{meyer2004hipass,meyer2017tracing}:
\begin{equation}
    S_{\rm \textsc{Hi}} = n \times \sigma_{\rm local} \times \sqrt{W \times \Delta v},
\end{equation}
where $\Delta v $ represents the spectral per-channel width  in $\text{km}/\text{s}$, and $n$ specifies the threshold ($n = 3$ for $3\sigma$ and $n = 5$ for $5\sigma$ limits).

Because the kinematic line widths are unknown for non-detected sources, we inferred the expected rotational velocities ($v_{\rm rot}$) using the empirical stellar mass Tully–Fisher relation \citep[TFR; e.g.][]{bell2001stellar,pizagno2007tully,lelli2016small}:
\begin{equation}
    \log_{10}\left(\frac{v_{\rm rot}}{\text{km}\,\text{s}^{-1}}\right) = \frac{\log_{10}(M_\star) - 1.9}{3.24},
    \label{eq:tfr}
\end{equation}
derived from the available stellar masses ($M_\star$). The full velocity width was subsequently estimated as $W \sim 2 \times v_{\rm rot}$, bounded to a physically motivated range of $50 \leq W \leq 400\,\text{km}\,\text{s}^{-1}$ to eliminate unphysical extremes for high- or low-mass systems. 

Finally, the upper limits on the \textsc{Hi} mass were calculated using the standard conversion, i.e., equation~\ref{eq:HI_mass_equ}. We note important caveats inherent to this approach. First, estimating $W$ via Equation~\ref{eq:tfr} assumes a uniform line profile and does not incorporate individual galaxy inclinations, meaning highly face-on systems may have over-predicted line widths. Second, the standard TFR exhibits a larger intrinsic scatter and a known breakdown in scaling at the low-mass end \citep[$M_\star \lesssim 10^8\,\text{M}_\odot$;][]{mcgaugh2012baryonic, lelli2016small, sales2017low}. 

\subsubsection{\rm \textbf{Complete \textsc{Hi} detection sample}}
\label{subsec:hi_uppr_limits}

To define a common \textsc{Hi}-complete detected sample, we adopt the sensitivity limit of the cluster observations, which are slightly less sensitive than the filament observations. This provides a common \textsc{Hi} mass range within which the detected galaxies in the two environments can be compared consistently.

At the average distance of the SSC ($D_{\rm L} \sim 210\,\text{Mpc}$, $z \sim 0.048$), the minimum detectable \textsc{Hi} mass at the pointing center — assuming a $3\sigma$ detection threshold integrated over a typical velocity width of $\sim 44\,\text{km}/\text{s}$ (i.e., per-channel spacing/width) and a central rms noise of $\sim 0.22\,\text{mJy}\,\text{beam}^{-1}$ — corresponds to $M_{\textsc{Hi}}^{\mathrm{center}} \sim 3.02 \times 10^8 \, \mathrm{M}_\odot$. This mass limit equates to an unattenuated column density sensitivity of $N_{\textsc{Hi}}^{\mathrm{center}} = 5.66 \times 10^{19} \, \text{cm}^{-2}$. However, due to the primary beam attenuation of the MeerKAT L-band system (which exhibits a FWHM$_{\rm PB}$ of $1.2^\circ$ at $1.28\,\text{GHz}$), the effective local sensitivity degrades rapidly as a function of angular distance from the pointing center.

At the conservative edge of our FoV, the primary beam response drops to $\sim 0.26$, translating to a flux attenuation factor of approximately $3.8$. Consequently, the local empirical detection limit at the spatial edge increases to $M_{\textsc{Hi}}^{\mathrm{edge}} \sim 1.15 \times 10^9 \, \mathrm{M}_\odot$ ($N_{\textsc{Hi}}^{\mathrm{edge}} = 2.16 \times 10^{20} \, \text{cm}^{-2}$) under identical extraction criteria. This implies that our overall detected sample within the supercluster is complete down to a global limit of $\log_{10}(M_{\textsc{Hi}}/\mathrm{M}_{\odot}) \sim 9$, though faint detections extend down to $\log_{10}(M_{\textsc{Hi}}/\mathrm{M}_{\odot}) \sim 8.5$ in the high-sensitivity inner regions of the FoV. 

\begin{figure}
	
	\includegraphics[angle=0,width=\columnwidth]{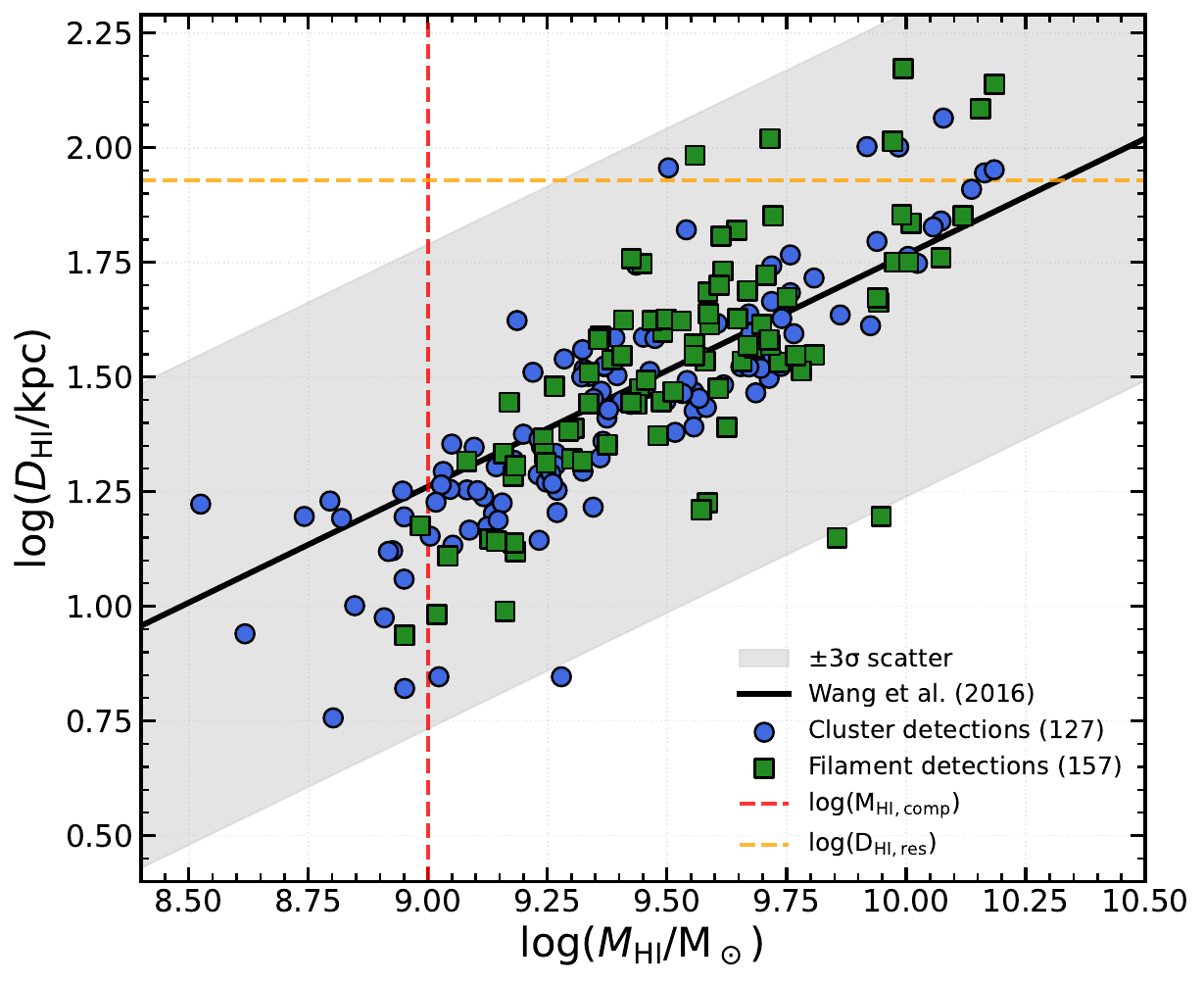}
    \caption{The \textsc{Hi} size--mass relation for the detected galaxy sample within the cluster centre (blue circles) and surrounding filamentary structure (green squares). The solid black line marks the reference empirical relation from \citet{wang2016new}, with the shaded grey region indicating the corresponding $\pm3\sigma$ intrinsic scatter. The horizontal orange dashed line outlines the spatial resolution limit, $\log_{10}(D_{\textsc{Hi}, \rm res}\rm/ kpc) \sim 1.93$, corresponding to three times the synthesized beam size ($3\theta_{\rm min}$). The vertical red dashed line indicates our \textsc{Hi} sample completeness limit $\log_{10}(M_{\textsc{Hi}, \rm comp}/\rm \text{M}_\odot) \sim 9$.} 
    \label{fig:hi_size_mass}
\end{figure}

This global completeness limit, $\log_{10}(M_{\textsc{Hi},\rm comp}/\mathrm{M}_\odot) \sim 9$, is highlighted by the red vertical dashed line in the size--mass distribution presented in Fig.~\ref{fig:hi_size_mass}. Using this threshold, we find that 113 out of 127 ($89$ per cent) cluster detections and 153 out of 157 ($97.5$ per cent) filament detections belong to this complete sample. For more details on how we determined the complete \textsc{Hi} detection sample of SSC, the reader is referred to \citet{gwebushe2026neutral}.

\subsubsection{\rm \textbf{\textsc{Hi} Diameter Determination}}
\label{subsec:hi_diameter}

To quantify the spatial extent of the \textsc{Hi} gas discs, we measure the \textsc{Hi} diameter ($D_{\textsc{Hi}}$) defined at the conventional surface density threshold of $\Sigma_{\textsc{Hi}} = 1\,\text{M}_\odot\,\text{pc}^{-2}$ \citep{broeils1997short,wang2016new}. 

First, the \textsc{Hi} total intensity (moment-0) map, $I_{\textsc{Hi}}$, is converted into an intrinsic \textsc{Hi} mass surface density map, $\Sigma_{\textsc{Hi}}$ ($\text{M}_\odot\,\text{pc}^{-2}$), following the conversion factor \citep[][]{meyer2017tracing}: 

\begin{equation}
    \Sigma_{\textsc{Hi}} = \frac{12.1}{\theta_{\rm maj} \times \theta_{\rm min}} I_{\textsc{Hi}},
\end{equation}

where $\theta_{\rm maj}$ and $\theta_{\rm min}$ represent the major and minor axes of the synthesized beam in arcseconds, respectively. 

The structural parameters of the \textsc{Hi} disc are determined by fitting a two-dimensional, rotated Gaussian profile to the spatial distribution of the moment-0 map via non-linear least-squares optimization. The fit yields the spatial deviations along the major and minor axes ($\sigma_x$ and $\sigma_y$). The apparent axis ratio is defined as $q = \min(\sigma_x, \sigma_y) / \max(\sigma_x, \sigma_y)$, bounded within $[0.05, 0.95]$ to minimize edge effects. Assuming an intrinsically thin disc geometry, the inclination angle $i$ is calculated via the geometric relation \citep{hubble1926extragalactic}:

\begin{equation}
    i = \arccos(q).
\end{equation}

The center of the \textsc{Hi} distribution is defined by computing the center of mass (intensity-weighted centroid) of the $\Sigma_{\textsc{Hi}}$ map. We isolate all pixels meeting or exceeding the threshold ($\Sigma_{\textsc{Hi}} \geq 1 \,\text{M}_\odot\,\text{pc}^{-2}$) and identify the maximum radial extent in pixels ($r_{\rm pix}$). 

We account for inclination projection effects by evaluating the radial extent along the unprojected major axis of the disc geometry, which provides the true physical radius of the system without requiring an additional $\cos(i)$ correction factor. Assuming the galaxy’s \textsc{Hi} distribution at the surface density is symmetric, the uncorrected \textsc{Hi} disc diameter is defined as $d_{\textsc{Hi},0} = 2r_{\textsc{Hi}}$. Assuming both the synthesized beam and the intrinsic \textsc{Hi} disc can be approximated as two-dimensional Gaussians, the beam-deconvolved \textsc{Hi} diameter ($D_{\textsc{Hi}}$) is derived using:
\begin{equation}
    D_{\textsc{Hi}} = \sqrt{d_{\textsc{Hi,0}}^2 - ab},
\end{equation}
where $a \equiv \theta_{\rm maj}$ and $b \equiv \theta_{\rm min}$ represent the synthesized beam major and minor axes, respectively. \textsc{Hi} diameters can only be reliably measured for galaxies that are spatially resolved. We classify galaxies with $D_{\textsc{Hi}} < 3 \times \theta_{\rm min}$ as unresolved (i.e. requiring a galaxy to span at least three resolution elements across its diameter) and treat their derived $D_{\textsc{Hi}}$ measurements as upper limits.


We plot our detected sample onto the empirical \textsc{Hi} size–mass relation from \citet{wang2016new} in Fig.~\ref{fig:hi_size_mass}. Of the 127 cluster and 157 filament galaxies detected in \textsc{Hi}, only 6 (4.7 per cent) and 6 (3.8 per cent) sources have resolved diameters exceeding this threshold ($D_{\textsc{Hi}} \geq 3\theta_{\rm min}$), respectively. 
For galaxies not detected in \textsc{Hi}, we estimate the maximum possible spatial extent of their \textsc{Hi} discs by applying the \citet{wang2016new} scaling law:

\begin{equation}
    \log_{10}(D_{\textsc{Hi},\rm lim}/\rm kpc) = 0.506 \log_{10}( \it M_{\textsc{Hi},\rm lim}/\rm M_{\odot}) - 3.293,
\end{equation}

evaluating this relation at each source's corresponding \textsc{Hi} mass upper limit to estimate upper limit \textsc{Hi} disc diameters.

Because of this small sample of resolved \textsc{Hi} detected sources, we cannot reliably quantify the \textsc{Hi} disc truncation relative to the optical disc ($D_{\textsc{Hi}}/D_{\rm opt}$) across the different targeted environments. In order to properly quantify environmental processing within the complex SSC using $D_{\textsc{Hi}}/D_{\rm opt}$ and directly compare our results to literature results \citep[e.g. the WALLABY study by][]{reynolds2022wallaby}, spatial resolution limitations must be taken into account. The Hydra I cluster is approximately four times closer than the SSC. Resolving comparable small-scale \textsc{Hi} structures in the SSC therefore requires about four times higher spatial resolution than achieved in this work. We therefore defer a disc truncation analysis to future work.
\section{\ion{H}{i} in cluster and filament galaxies}
\label{sec:Results}

Due to varying observational and data constraints, some physical quantities cannot be uniformly determined for every galaxy within our sample. Consequently, the sample size varies slightly between analyses depending on the availability and quality of the required measurements.

For some galaxies, stellar masses could not be estimated because the optical photometry was affected by imaging artefacts or contamination from bright foreground stars. These galaxies are excluded from analyses requiring stellar masses, resulting in small variations in sample size between different analyses.

We also verified that our results are not sensitive to the adopted sample selection. Consistent with \citet{gwebushe2026neutral}, we find no significant differences between the parameter distributions obtained using the \textsc{Hi}-complete sample and those derived from the full \textsc{Hi}-detected sample. We therefore use the full sample throughout this work to maximise the statistical significance.

Finally, the \textsc{Hi} non-detections in the cluster centre and filament regions are treated as upper limits on their \textsc{Hi} masses. Although the cluster and filament observations have comparable rms noise levels, the corresponding upper limits also depend on the local sensitivity, primary-beam attenuation, and the assumed line width of each galaxy. We therefore retain the \textsc{Hi} non-detections as upper limits and show them where useful for illustrating the sensitivity limits of the observations. They are not treated as measured \textsc{Hi} masses, and all quantitative comparisons and statistical tests of \textsc{Hi}-derived properties are based on secure \textsc{Hi} detections. The non-detections are included only to indicate the upper-limit constraints and the population for which direct \textsc{Hi} mass measurements are not available.


\subsection{Scaling Relations and Parameter Offsets across Star-formation and Gas-fraction Sequences}
\label{sec:HI_cont_&_SF}

In this subsection, we investigate how the \textsc{Hi} gas content of galaxies in different supercluster environments relates to their star formation activity and stellar mass. To do this, we follow the analysis methods described by \citet{janowiecki2020xgass}. The breakdown of our galaxy sample across the different star-forming and gas-content classifications is summarized in Table~\ref{tab:sample_breakdown}.

\begin{figure*}
    \centering
    \includegraphics[width=\textwidth]{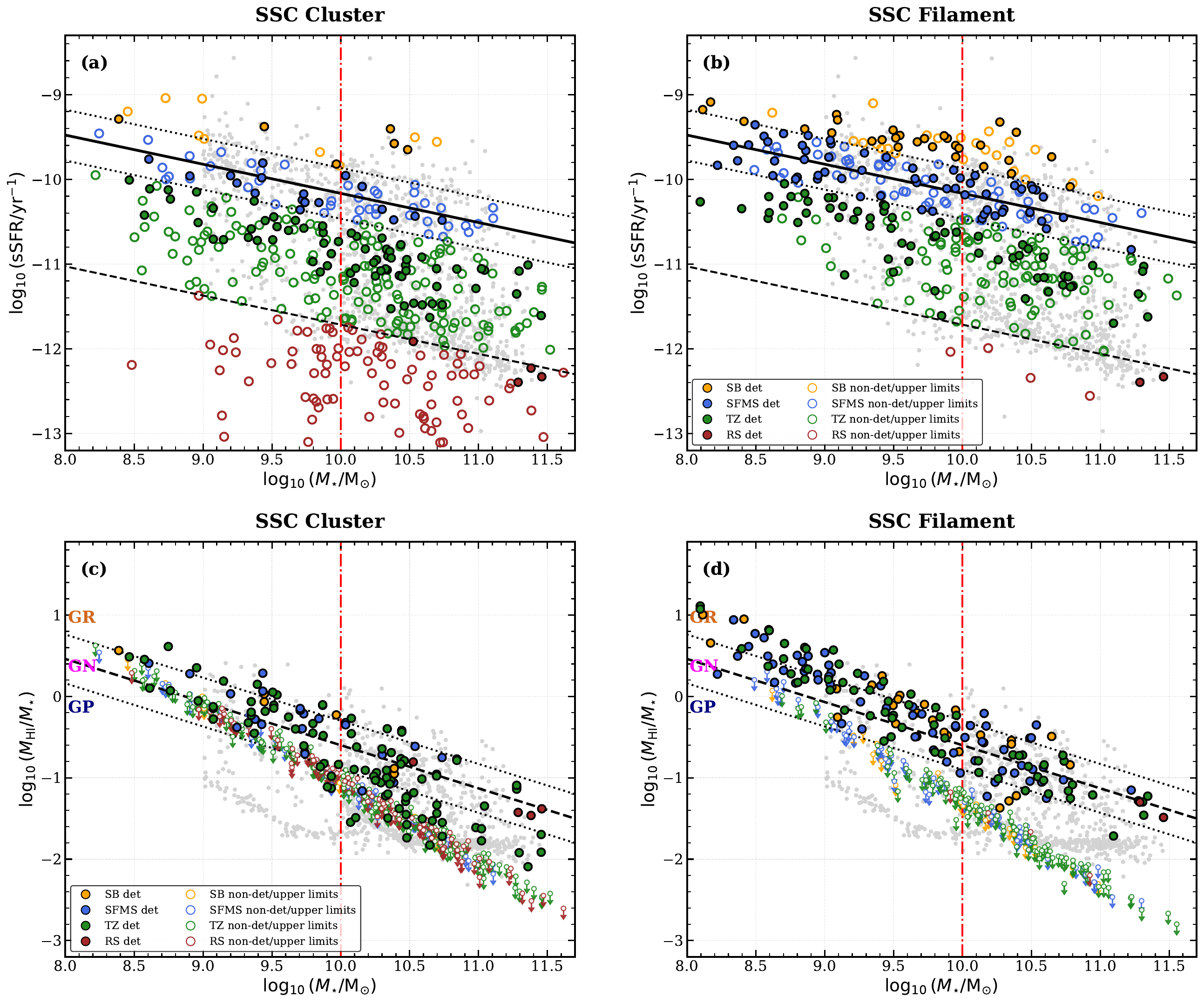}
    \caption{Overview of the star formation and \textsc{Hi} scaling relations for galaxies in the SSC compared to the field population. In all panels, the light grey background points show the reference field population from the xGASS survey \citep{janowiecki2020xgass}. Solid markers represent MeerKAT \textsc{Hi} detections, while open markers (or points with downward arrows) indicate \textsc{Hi} non-detections (upper limits). Sources are color-coded by their star-forming state: starbursts (SB; orange), star-forming main sequence (SFMS; blue), transition zone (TZ; green), and red sequence (RS; brown). The vertical red dot-dashed line marks the stellar mass threshold at $\log_{10}(M_{\star}/\text{M}_\odot) = 10$. 
    {\it Top Row (panels a and b):} Specific star formation rate ($\text{sSFR}$) as a function of stellar mass ($M_{\star}$) for the SSC cluster (left) and the surrounding SSC filament (right). The solid black line marks the main sequence, bounded by $\pm0.3\,\text{dex}$ scatter lines (dotted) and the quenching threshold (dashed line).
    {\it Bottom Row (panels c and d):} \textsc{Hi} gas mass fraction, $\log_{10}(M_{\textsc{Hi}}/M_{\star})$, as a function of stellar mass for the SSC cluster (left) and the filament (right). The dashed and dotted black lines indicate the median scaling relation and corresponding scatter from xGASS. Text labels in the top left corner of panels (c) and (d) mark the positions of Gas-Rich (GR), Gas-Normal (GN), and Gas-Poor (GP) populations.}
    \label{fig:sSFR_gasF_scaling_relation}
\end{figure*}

Fig.~\ref{fig:sSFR_gasF_scaling_relation} (top row) shows specific star formation rate ($\text{sSFR}$) plotted against stellar mass ($M_{\star}$) for cluster galaxies (panel a) and filament galaxies (panel b). 
We divide our galaxy sample into four star-forming sub-classes: starburst (SB), star-forming main sequence (SFMS), transition zone (TZ), and red sequence (RS). The solid black line represents the reference main sequence derived from the xGASS sample by \citet{janowiecki2020xgass}. The dotted black lines above and below this line represent a scatter of $\pm 0.3\,\text{dex}$ ($\sim 1\sigma$). Galaxies within this range are classified as SFMS systems, while those above it are classified as SB galaxies. To separate the remaining population, we follow \citet{janowiecki2020xgass} and use a horizontal offset threshold of $\Delta \text{SFMS} = -1.55\,\text{dex}$ (i.e. black dashes) to define the RS boundary. Galaxies that lie in between the SFMS and the RS boundaries are classified as TZ galaxies.

Looking at the star-forming populations in the cluster environment (panel a), the \textsc{Hi}-detected sample consists of $73.2$ per cent TZ galaxies, $3.1$ per cent RS galaxies, $17.3$ per cent SFMS galaxies, and $6.3$ per cent SB galaxies. In contrast, the cluster non-detected sample is dominated by passive systems, containing $50.7$ per cent TZ galaxies, $31.6$ per cent RS galaxies, $14.6$ per cent SFMS galaxies, and a small fraction ($3.1$ per cent) of SB galaxies. In the filament environment (panel b), the \textsc{Hi}-detected sample contains $40.1$ per cent TZ galaxies, $1.3$ per cent RS galaxies, $39.5$ per cent SFMS galaxies, and $11.5$ per cent SB galaxies. The filament non-detected sample consists of $55.0$ per cent TZ galaxies, $2.1$ per cent RS galaxies, $31.4$ per cent SFMS galaxies and $11.5$ per cent SB galaxies.

The bottom row of Fig.~\ref{fig:sSFR_gasF_scaling_relation} shows  \textsc{Hi} gas mass fraction, $\log_{10}(M_{\textsc{Hi}}/M_{\star})$, plotted against the stellar mass for the cluster (panel c) and filament (panel d) environments. Following \citet{janowiecki2020xgass}, we separate the galaxies into three categories: gas-rich (GR), gas-normal (GN), and gas-poor (GP). The GN population is defined as galaxies within $\pm 0.3\,\text{dex}$ (dotted lines) of the median gas fraction main sequence (dashed line). Galaxies above this region are classified as GR, and those below it are GP.

In the cluster environment (panel c), most of the \textsc{Hi}-detected galaxies have a normal gas content, with $60.6$ per cent classified as GN, $13.4$ per cent as GR, and $26.0$ per cent as GP. 
In the lower-density filament environment (panel d), the majority of the \textsc{Hi}-detected galaxies are GN ($63.1$ per cent), followed by GR ($31.2$ per cent) and GP ($5.7$ per cent). 

From Fig.~\ref{fig:sSFR_gasF_scaling_relation}, we can make a direct observation about the nature of the detected galaxies across both environments. While the cluster centre (panel a) is dominated by TZ systems, the filament sample (panel b) contains comparable fractions of SFMS and TZ galaxies. 
When we look at their gas content in the bottom panels, we find differences between the two environments. In the SSC cluster (panel c), most of the \textsc{Hi}-detected galaxies are classified as GN, indicating gas content close to the xGASS-defined median relation. In the SSC filament (panel d), the majority of the detected population is also classified as GN, followed by GR. The filament population contains a larger fraction of GR galaxies than the cluster population and this prominent population of GR systems in the filament suggests that the lower-density environment allows galaxies to retain their gas reservoirs more efficiently than those inside the dense cluster centre.

\begin{table*}
\centering
\caption[Sample breakdown of galaxies in the SSC cluster and SSC filament environments classified by star-forming state and \textsc{Hi} gas content category]{Sample breakdown of galaxies in the SSC Cluster and SSC Filament environments classified by star-forming state (top rows) and \textsc{Hi} gas content category (bottom rows). Percentages indicate the fraction relative to the total population in each respective column.}
\label{tab:sample_breakdown}
\begin{tabular}{lcccc}
\hline
\hline
\noalign{\smallskip}
 & \multicolumn{2}{c}{\textbf{SSC cluster}} & \multicolumn{2}{c}{\textbf{SSC filament}} \\
\textbf{Classification} & \textsc{Hi} Det & \textsc{Hi} Non-det & \textsc{Hi} Det & \textsc{Hi} Non-det \\
\noalign{\smallskip}
\hline
\noalign{\smallskip}\\
\multicolumn{5}{l}{\textbf{Star-Forming State}} \\
\noalign{\smallskip}
Starburst (SB)             &  6.3 per cent [8]    & 3.1 per cent [9]    & 19.1 per cent [30]   & 11.5 per cent [22]   \\
Main Sequence (SFMS)       & 17.3 per cent [22]  & 14.6 per cent [43]   & 39.5 per cent [62] & 31.4 per cent [60]  \\
Transition Zone (TZ)       & 73.2 per cent [93]  & 50.7 per cent [149] & 40.1 per cent [63] & 55.0 per cent [105]\\
Red Sequence (RS)          & 3.1 per cent [4]  & 31.6 per cent [93] & 1.3 per cent [2] & 2.1 per cent [4]\\
\noalign{\smallskip}
\textbf{Total}             & \textbf{[127]} & \textbf{[294]} & \textbf{[157]} & \textbf{[191]} \\
\noalign{\smallskip}
\hline
\noalign{\smallskip}\\
\multicolumn{5}{l}{\textbf{Gas Content Category}} \\
\noalign{\smallskip}
Gas-Rich (GR)              & 13.4 per cent [17]  & 0.0 per cent [0]    & 31.2 per cent [49] & 0.0 per cent [0]   \\
Gas-Normal (GN)            & 60.6 per cent [77]  & 40.9 per cent [93]  & 63.1 per cent [99] & 11.5 per cent [22]   \\
Gas-Poor (GP)              & 26.0 per cent [33]  & 58.7 per cent [201] & 5.7 per cent [9]   & 88.5 per cent [169]\\
\noalign{\smallskip}
\textbf{Total}             & \textbf{[127]} & \textbf{[294]} & \textbf{[157]} & \textbf{[191]} \\
\noalign{\smallskip}
\hline
\hline
\end{tabular}
\end{table*}


\begin{figure*}
    \centering
    \includegraphics[width=\textwidth]{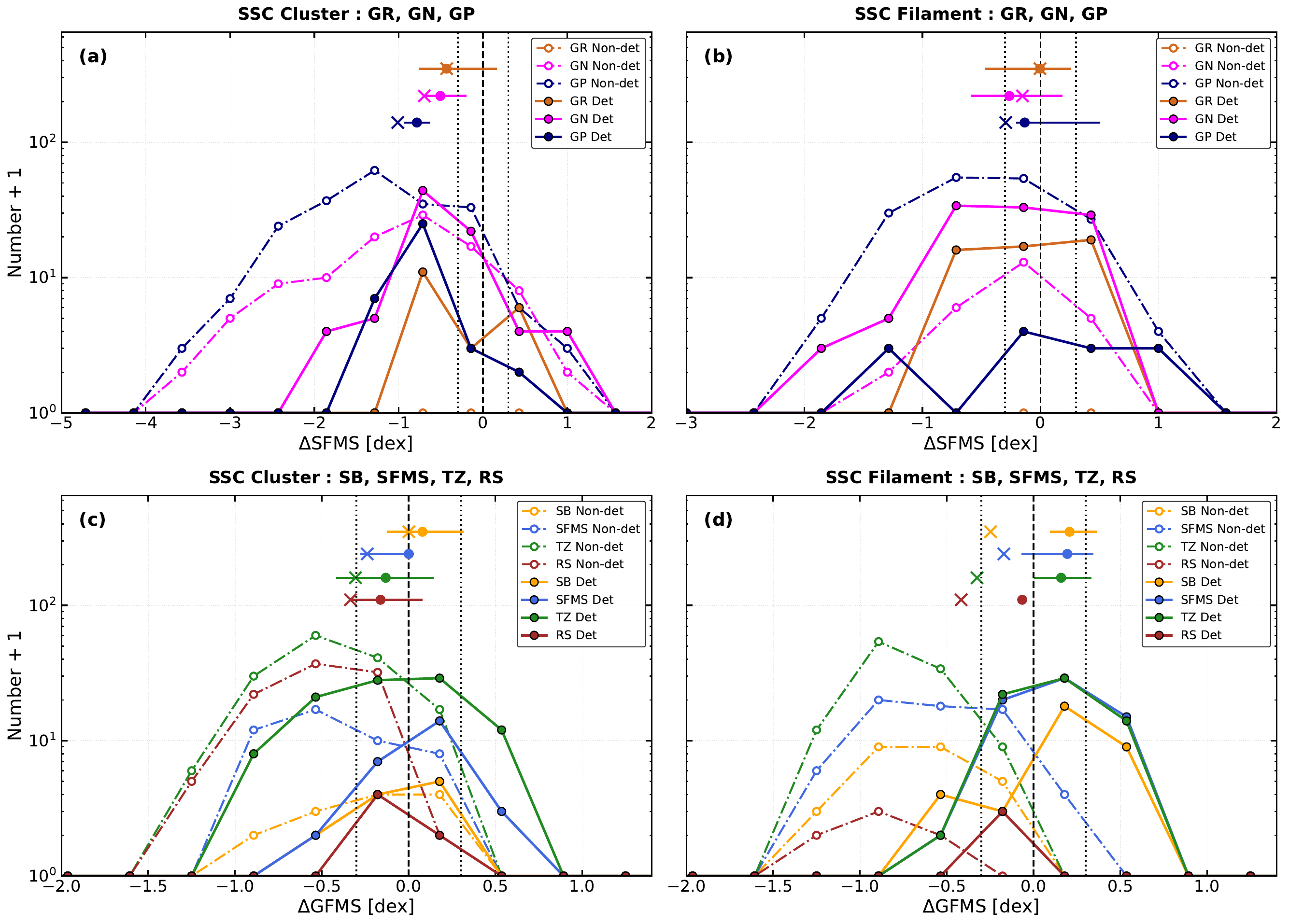}
    \caption{Distributions of scaling relation offsets within the SSC cluster centre and SSC filament environments. The top row shows the number distribution of galaxies as a function of their vertical displacement from the star-forming main sequence ($\Delta \mathrm{SFMS}$), color-coded by gas mass fraction sub-populations: Gas-Rich (GR, chocolate), Gas-Normal (GN, magenta), and Gas-Poor (GP, navy). The bottom row displays the number distribution as a function of vertical displacement from the gas fraction main sequence ($\Delta \mathrm{GFMS}$), split into starburst (SB, orange), star-forming main sequence (SFMS, royal blue), transition zone (TZ, forest green), and red sequence (RS, brown) sub-classes. In all panels, circles indicate the median positions of the $\textsc{Hi}$-detected sub-samples with horizontal error bars showing the 25th to 75th percentile ranges, while crosses label the mean of the individual detected and non-detected medians.}
    \label{fig:SFMS_GFMS_Staggered_Aligned}
\end{figure*}

To investigate these environmental variations in more detail, we look at how the vertical offsets from our scaling relations correlate with galaxy sub-populations. We first consider the distance of galaxies above and below the star-forming main sequence, defined as $\Delta \text{SFMS} \equiv \log_{10}(\text{sSFR}) - \log_{10}(\text{sSFR}_{\text{MS}})$ \citep[][]{janowiecki2020xgass}. The top panels of Fig.~\ref{fig:SFMS_GFMS_Staggered_Aligned} present the distribution of $\Delta \text{SFMS}$ binned by gas fraction categories (GR, GN, and GP) for both the cluster (panel a) and filament environment (panel b). 

We also analyze the vertical displacement of galaxies relative to the gas fraction main sequence, defined as $\Delta \text{GFMS} \equiv \log_{10}(f_ {\rm \textsc{Hi}}) - \log_{10}(f_{\rm \textsc{Hi,MS}})$ \citep[][]{janowiecki2020xgass}. The bottom panels of Fig.~\ref{fig:SFMS_GFMS_Staggered_Aligned} display the distribution of $\Delta \text{GFMS}$ separated by star-forming sub-classes (SB, SFMS, TZ, and RS) for the cluster (panel c) and filament (panel d). The median values alongside the 25th and 75th percentile ranges for the $\textsc{Hi}$-detected galaxies (circles) and the mean of the medians combining the detections and non-detections (crosses) are annotated within the plot layout (i.e. above the distributions). These statistical constraints are summarized in Table~\ref{tab:offset_statistics}.

\begin{table*}
\centering
\caption{Statistical summary of the vertical offsets ($\Delta \text{SFMS}$ and $\Delta \text{GFMS}$) across the galaxy sub-populations in the SSC cluster and SSC filament environments. Median values and error bounds representing the 25th and 75th percentiles are shown for the $\textsc{Hi}$-detected samples, alongside the mean of the medians calculated between the detected and non-detected populations.}
\label{tab:offset_statistics}
\begin{tabular}{lcccccc}
\hline
\hline
\noalign{\smallskip}
 & \multicolumn{3}{c}{\textbf{SSC cluster}} & \multicolumn{3}{c}{\textbf{SSC filament}} \\
 \cline{2-4} \cline{5-7}
\noalign{\smallskip}
\textbf{Sub-population} & Med (Det) & 25th / 75th Pctl & Mean of Med & Med (Det) & 25th / 75th Pctl & Mean of Med \\
\noalign{\smallskip}
\hline
\noalign{\smallskip}\\
\multicolumn{7}{l}{\rm \textbf{\textup{$\Delta \text{SFMS}$ Distributions [dex]}}} \\
Gas-Rich (GR)   & $-0.43$ & $[-0.76, 0.16]$ & $-0.43$ & $-0.005$ & $[-0.47, 0.26]$ & $-0.005$ \\
Gas-Normal (GN) & $-0.51$ & $[-0.70, -0.20]$ & $-0.70$ & $-0.26$ & $[-0.59, 0.19]$ & $-0.15$ \\
Gas-Poor (GP)   & $-0.79$ & $[-0.94, -0.62]$ & $-1.01$ & $-0.13$ & $[-0.21, 0.51]$ & $-0.29$ \\
\noalign{\smallskip}
\hline
\noalign{\smallskip}\\
\multicolumn{7}{l}{\textup{\textbf{$\Delta \text{GFMS}$ Distributions [dex]}}} \\
Starburst (SB)       & $0.08 $   & $[-0.13, 0.32]$ & $-0.001$ & $0.21$  & $[0.09, 0.37]$  & $-0.28$  \\
Main Sequence (SFMS) & $-0.001$  & $[-0.27, 0.001]$  & $-0.24$ & $0.19$  & $[-0.07, 0.35]$  & $-0.21$ \\
Transition Zone (TZ) & $-0.13$ & $[-0.42, 0.14]$ & $-0.30$ & $0.15$  & $[0.003, 0.33]$  & $-0.31$ \\
Red Sequence (RS)    & $-0.16$ & $[0.08, 0.002]$ & $-0.43$ & $-0.07$  & $[-0.08, -0.04]$  & $-0.39$ \\
\noalign{\smallskip}
\hline
\hline
\end{tabular}
\end{table*}

From Fig.~\ref{fig:SFMS_GFMS_Staggered_Aligned} and Table~\ref{tab:offset_statistics}, several behaviors can be noted regarding the progression of quenching. In panel (a), all gas sub-classes (GR, GN, and GP) in the cluster environment exhibit deep negative median offsets below the main sequence, between $\sim -0.79\,\text{dex}$ and $\sim -0.43\,\text{dex}$. This shows that this sample is dominated by passive, low-star-forming systems. In panel (b), the gas sub-classes exhibit smaller negative median offsets, all within the $\pm0.3$ dex scatter of the SFMS. This suggests that the filament population retains relatively high star formation activity, although it also includes galaxies with reduced star formation rates. However, the relatively low detection fractions of the GR, GN, and GP sub-samples limit the statistical significance and robustness of these distributions.



We assess the significance of difference in the distributions using the Kolmogorov–Smirnov (KS) test, where the difference between the distributions is statistically significant for $p$-values $<0.05$. For $p$-values $\geq0.05$, we fail to reject the hypothesis that the samples come from the same distribution. The KS statistic, $D$, is defined as the maximum absolute difference between the empirical cumulative distribution functions of any two samples. 
Larger values of $D$ indicate a greater difference between the distributions.

Given the relatively large number of KS tests performed in this analysis ($\sim 30$), the statistical significance of individual comparisons should be interpreted with caution. When multiple hypothesis tests are conducted at a significance level of 5 per cent, one or two statistically significant results may be expected by chance alone, even if the underlying distributions are drawn from the same parent population. Therefore, the significant differences identified in individual comparisons, particularly those with p-values close to 0.05, should be regarded as indicative rather than conclusive evidence of differences between the populations. These results are interpreted alongside the observed distributions and median values rather than in isolation.

Within the cluster centre (panel a), the KS test shows that the star formation distributions of the GR and GN populations are statistically indistinguishable ($D = 0.22$, $p \approx 0.46$). However, the GP population represents a distinct underlying distribution, showing significant variations when compared against both the GR population ($D = 0.50$, $p < 0.004$) and the GN population ($D = 0.47$, $p \approx 0.001$). 

In contrast, within the lower-density filamentary environment (panel b), no statistically significant differences are detected among any of the gas sub-populations. The KS test yields no significant variation between the filament GR and GN distributions ($D = 0.17$, $p \approx 0.24$), nor when comparing the gas-poor tracks to the GR ($D = 0.25$, $p \approx 0.62$) and GN ($D = 0.31$, $p \approx 0.34$) components. This indicates a more uniform distribution of star formation across gas classes outside the cluster centre. 

The KS tests reveal statistically significant differences in the star formation distributions between the cluster and filament environments for all three gas sub-populations. The comparisons yield $D = 0.38$ ($p \approx 0.04$) for GR, $D = 0.30$ ($p \approx 0.001$) for GN, and $D = 0.72$ ($p \approx 0.001$) for GP galaxies. These results suggest that galaxies with similar \textsc{Hi} gas content exhibit different star formation properties depending on their environment, highlighting the role of environment in shaping the star formation activity of galaxies.


In the cluster centre (panel c), a clear \textsc{Hi} depletion trend is visible across the star formation populations. Detected SB galaxies exhibit a gas excess, with a median \(\Delta\mathrm{GFMS}\sim 0.08\) dex, while SFMS galaxies remain GN, with a median $\Delta\mathrm{GFMS} \sim -0.001$ dex. However, as galaxies move below the SFMS, the median \textsc{Hi} gas fraction offset becomes progressively more negative, although the values remain within the \(\pm 0.3\) dex scatter of the GFMS. The cluster TZ population shows a negative offset ($\text{median} \sim -0.13\,\text{dex}$), which continues to drop until it reaches a minimum in the RS state, which exhibits the lowest median ($\text{median} \sim -0.16\,\text{dex}$).

In contrast, the filament environment (panel d) shows a positive gas-fraction offset across all star-forming sub-classes, except for RS with negligible number of \textsc{Hi} detection. SB filament galaxies show the highest gas excess, with a median $\Delta\mathrm{GFMS} \sim 0.21\,\mathrm{dex}$. This positive offset persists for galaxies on the SFMS ($\mathrm{median}\sim0.19\,\mathrm{dex}$) and through the filament TZ population ($\mathrm{median}\sim0.15\,\mathrm{dex}$). The small sample of RS galaxies show negative offset, with a median $\Delta\mathrm{GFMS}\sim-0.07\,\mathrm{dex}$. 

Within the cluster centre, the $\Delta\mathrm{GFMS}$ distributions for SB galaxies does not show any significant difference with other populations, with SFMS ($D = 0.28$, $p \approx 0.63$), with TZ ($D = 0.37$, $p \approx 0.21$), with RS ($D = 0.50$, $p \approx 0.51$). The $\Delta\mathrm{GFMS}$ distributions differ significantly between the SFMS and TZ populations ($D = 0.41$, $p \approx 0.004$), while the TZ and RS populations do not show a statistically significant difference ($D = 0.49$, $p \approx 0.21$). The SFMS and RS populations also show no statistically different distributions ($D = 0.59$, $p \approx 0.13$). 

In the filament environment, none of the comparisons between classes reveal statistically significant differences in their \(\Delta\mathrm{GFMS}\) distributions. The SB population shows no significant difference from the SFMS (\(D = 0.18\), \(p \approx 0.44\)), TZ (\(D = 0.26\), \(p \approx 0.12\)), or RS populations (\(D = 0.83\), \(p \approx 0.08\)). Similarly, no significant differences are found between the SFMS and TZ (\(D = 0.10\), \(p \approx 0.85\)), TZ and RS (\(D = 0.71\), \(p \approx 0.18\)), or SFMS and RS (\(D = 0.68\), \(p \approx 0.23\)) populations. Thus, despite the differences in the median offsets, the \(\Delta\mathrm{GFMS}\) distributions of the \textsc{Hi}-detected filament galaxies are statistically consistent across the evolutionary classes. This suggests no clear evidence for a systematic decline in \textsc{Hi} gas fraction towards the RS population within the \textsc{Hi}-detected filament sample. This result applies to the \textsc{Hi}-detected population and does not imply that the full filament galaxy population has the same \textsc{Hi} gas-fraction distribution.

Finally, the cross-environment KS tests show that the \(\Delta\mathrm{GFMS}\) distributions are statistically consistent between the cluster centre and filaments for the SB (\(D = 0.33\), \(p \approx 0.41\)), SFMS (\(D = 0.26\), \(p \approx 0.17\)), and RS (\(D = 0.50\), \(p \approx 0.93\)) populations. A statistically significant difference is found only for the TZ population (\(D = 0.35\), \(p \approx 0.001\)). 


\subsection{Mass-Dependent Offset Profiles: Mass-Driven vs. Environmental Signatures}
\label{sec:HI_cont_&_M*}

\begin{figure*}
\centering
\includegraphics[width=\textwidth]{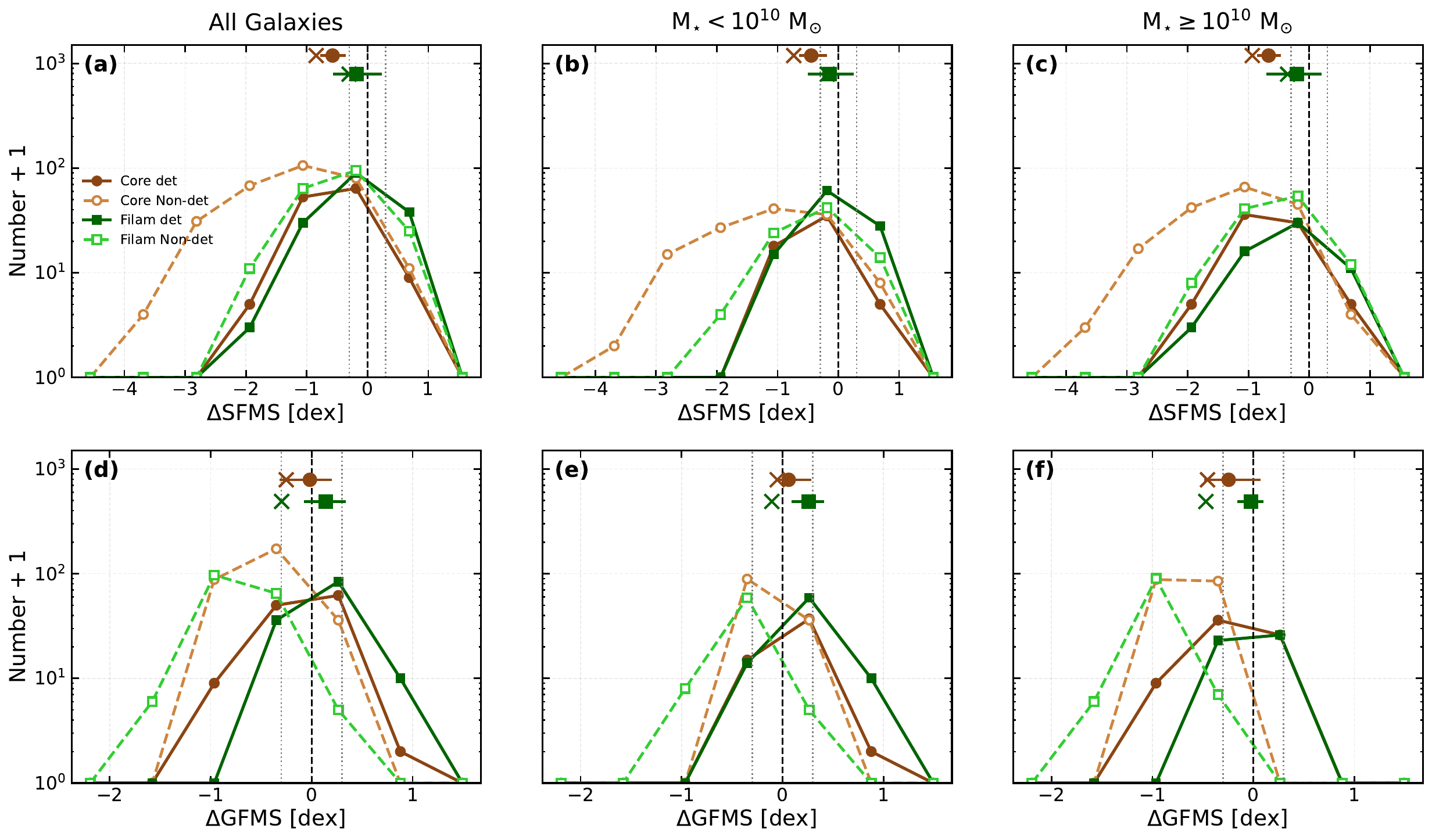}
\caption{Distributions of the scaling relation offsets across different environmental and stellar mass regimes. The top row (panels a--c) presents the histograms for the star-forming main sequence offsets ($\Delta\text{SFMS}$), while the bottom row (panels d--f) details the gas fraction main sequence offsets ($\Delta\text{GFMS}$). The left, center, and right columns represent the total sample, the low-mass subsample ($M_{\star} < 10^{10}\,\text{M}_{\odot}$), and the high-mass subsample ($M_{\star} \geq 10^{10}\,\text{M}_{\odot}$), respectively. Markers above the histograms display the median and 25th/75th percentile bounds for the $\textsc{Hi}$-detected cluster centre (brown circles) and filament (green squares). Crosses denote the mean of the medians evaluated across the joint detected and non-detected components. Dashed lines mark the zero-offsets, with dotted boundaries reporting the $\pm0.3\,\text{dex}$ scatter envelope.}
\label{fig:SFMS_GFMS_histograms}
\end{figure*}

The top and bottom rows of Fig.~\ref{fig:SFMS_GFMS_histograms} show histograms of the vertical offset from $\Delta\text{SFMS}$ and the offset from $\Delta\text{GFMS}$, respectively. The median alongside the 25th and 75th percentiles of each distribution is also plotted to illustrate the inner-quartile ranges of the $\textsc{Hi}$-detected galaxies. The mean of the median position of the combined $\textsc{Hi}$ detections and non-detections is marked by crosses. 

The left columns of Fig.~\ref{fig:SFMS_GFMS_histograms} show the full galaxy population within each environment. To check if the trends present in the full sample are driven or modified by stellar mass, we subdivide the galaxy populations into low- and high-stellar-mass subsamples ($M_{\star} < 10^{10}\,\text{M}_{\odot}$ and $M_{\star} \geq 10^{10}\,\text{M}_{\odot}$, marking the boundary between low and high mass systems) in the center and right columns. 
The statistical properties of these distributions are summarised in Table~\ref{tab:hist_statistics}.

\begin{table*}
\centering
\caption{Statistical summary of the vertical offsets ($\Delta \text{SFMS}$ and $\Delta \text{GFMS}$) binned by environment and stellar mass bins. Median values and $25\text{th}/75\text{th}$ percentile error bounds are given for the $\textsc{Hi}$-detected samples. The mean of the medians integrates both the detected and non-detected tracking profiles.}
\label{tab:hist_statistics}
\begin{tabular}{lcccccccc}
\hline
\hline
\noalign{\smallskip}
 & \multicolumn{4}{c}{\textbf{SSC cluster}} & \multicolumn{4}{c}{\textbf{SSC filament}} \\
\cline{2-5} \cline{6-9}
\noalign{\smallskip}
\textbf{Mass Slice} & $N_{\text{det}}$ & Med (Det) & 25th / 75th Pctl & Mean of Med & $N_{\text{det}}$ & Med (Det) & 25th / 75th Pctl & Mean of Med \\
\noalign{\smallskip}
\hline
\noalign{\smallskip}\\
\multicolumn{9}{l}{\textbf{\textup{$\Delta \text{SFMS}$ Distributions [dex]}}} \\
Full Sample & 127 & $-0.57$ & $ [-0.78, -0.36]$ & $-0.84$ & 157 & $-0.18$ & $[-0.56, 0.23]$ & $-0.30$ \\
$M_\star < 10^{10}\,\text{M}_{\odot}$ & 55 & $-0.45$ & $[-0.64, -0.18]$ & $-0.73$ & 101 & $-0.15$ & $[-0.50,0.26]$ & $-0.18$ \\
$M_\star \geq 10^{10}\,\text{M}_{\odot}$ & 72 & $-0.67$ & $[-0.86, -0.47]$ & $-0.94$ & 56  & $-0.20$ & $[-0.70, 0.21]$ & $-0.36$ \\
\noalign{\smallskip}
\hline
\noalign{\smallskip}\\
\multicolumn{9}{l}{\textbf{\textup{$\Delta \text{GFMS}$ Distributions [dex]}}} \\
Full Sample & 127 & $-0.02$ & $[-0.32, 0.20]$ & $-0.25$ & 157 & $0.14$ & $[-0.08, 0.34]$ & $-0.30$ \\
$M_\star < 10^{10}\,\text{M}_{\odot}$ & 55 & $0.06$ & $[-0.06, 0.28]$ & $-0.05$ & 101 & $0.26$ & $[0.09, 0.41]$ & $-0.11$ \\
$M_\star \geq 10^{10}\,\text{M}_{\odot}$ & 72 & $-0.24$ & $[-0.49, 0.07]$ & $-0.45$ & 56 & $-0.03$ & $[-0.16, 0.10]$ & $-0.47$ \\
\noalign{\smallskip}
\hline
\hline
\end{tabular}
\end{table*}

Concerning the $\Delta\text{SFMS}$ profiles (top row), we see that moving from the low-mass to the high-mass subsamples, the star formation shifts towards  negative values for both subsamples. For instance, the cluster centre detection medians drop from $-0.45\,\text{dex}$ to $-0.67\,\text{dex}$, for the filament detections, the shift goes from $-0.15\,\text{dex}$ to $-0.20\,\text{dex}$.

The cross-environment KS tests show no significant differences between the cluster centre and filament detections for the full sample ($D = 0.07$, $p \approx 0.76$), the low-mass systems ($D = 0.14$, $p \approx 0.33$), or the high-mass systems ($D = 0.10$, $p \approx 0.91$). However, the underlying non-detected gas reservoirs reveal a different trend; the non-detection distributions vary significantly across environment for the full sample ($p < 0.001$) and the low-mass bin ($p < 0.001$), demonstrating that environmental modifications impact the non-detected, gas-poor field before altering the active $\textsc{Hi}$ core signatures.

The bottom row ($\Delta\text{GFMS}$) reveals a significant divergence in \textsc{Hi} gas content between the two environments. In the cluster centre, the full sample of detections lies close to zero ($\text{median} = -0.02\,\text{dex}$). When looking at the mass bins, low-mass cluster detections retain a small gas excess ($\text{median} = 0.06\,\text{dex}$), whereas high-mass systems dip into a negative offset ($\text{median} = -0.24\,\text{dex}$). Because \textsc{Hi} detectability depends on both the \textsc{Hi} mass and the physical properties of the galaxy, including its stellar mass and expected line width, the mass-dependent distributions shown here describe the \textsc{Hi}-detected population and should not be interpreted as representative of the full galaxy population at each stellar mass.

Filament detections show a positive gas offset, peaking at $0.26\,\text{dex}$ for the low-mass systems and decreasing to $-0.03\,\text{dex}$ in the high-mass bin. 
The cross-environment KS tests show that the underlying gas fraction distributions for matching mass categories are distinct between the cluster centre and filament for the total population ($D = 0.29$, $p < 0.001$), the low-mass systems ($D = 0.27$, $p < 0.02$), and the high-mass systems ($D = 0.38$, $p < 0.001$). These statistically significant differences indicate that the large-scale environment influences the \textsc{Hi} gas content of galaxies. While galaxies in the high-density cluster centre exhibit progressively lower \textsc{Hi} gas fractions, filament galaxies retain relatively high \textsc{Hi} gas fractions. The connection between environmental effects, \textsc{Hi} content, and star formation in intermediate-density environments is consistent with early observations of cluster and supercluster fields \citep[e.g.][]{haynes1984neutral,giovanelli1985gas}, as well as studies of the Coma Supercluster \citep[e.g.][]{gavazzi2010snapshot}, which show that star formation can be suppressed before complete removal of the \textsc{Hi} reservoir during environmental processing.

\subsection{Gas Depletion Timescales Across cluster and filament Environments}

The gas depletion timescale, defined as $t_{\mathrm{dep}} \equiv M_{\mathrm{gas}}/\mathrm{SFR}$, represents the time a galaxy would take to exhaust its current cold gas reservoir at its ongoing star formation rate.  Since it is the inverse of the star formation efficiency (SFE), it provides insight into how efficiently a galaxy is forming stars. A short $t_{\mathrm{dep}}$ corresponds to high SFE, making $t_{\mathrm{dep}}$ a key diagnostic for both star formation activity and quenching processes \citep[e.g.][]{saintonge2011cold, tacconi2018phibss}. Short $t_{\mathrm{dep}}$ are typically associated with SB galaxies that are rapidly consuming their gas, while longer $t_{\mathrm{dep}}$ can either reflect inefficient star formation or galaxies in the early stages of quenching \citep{catinella2018xgass}.

In dense environments such as galaxy clusters, $t_{\mathrm{dep}}$ is affected also by environmental quenching processes—such as ram pressure stripping, starvation (strangulation), and tidal interactions—on both star formation activity and gas content \citep{boselli2022ram, janowiecki2020xgass, brown2023vertico}. For example, galaxies experiencing rapid gas removal may exhibit short $t_{\mathrm{dep}}$ due to active star formation with limited remaining gas, while galaxies undergoing starvation may retain their \textsc{Hi} reservoirs but show suppressed star formation, leading to longer $t_{\mathrm{dep}}$ values. 

\begin{figure*}
\centering
\includegraphics[width=\textwidth]{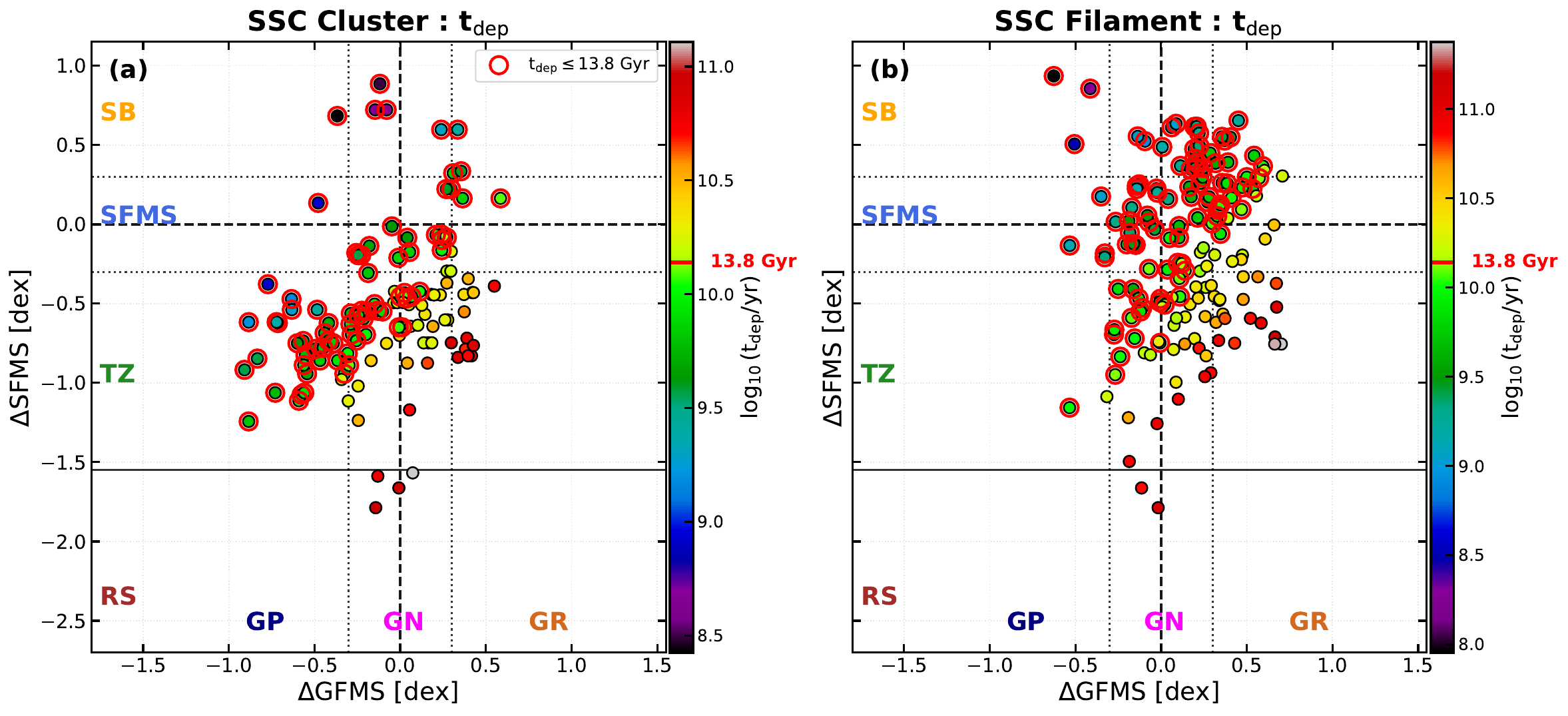}
\caption{Vertical offset from the star-forming main sequence ($\Delta\text{SFMS}$) as a function of the gas fraction main sequence offset ($\Delta\text{GFMS}$) for the detected SSC cluster centre (panel a) and the SSC filaments (panel b). Individual galaxies are color-coded by their gas depletion timescale, $\log_{10}(t_{\mathrm{dep}}/\text{yr})$. 
Red open rings emphasize systems with efficient or rapid consumption tracks, $t_{\mathrm{dep}} \leq 13.8\,\text{Gyr}$, and a horizontal solid red mark on the colorbar isolates the Hubble time threshold, $13.8\,\text{Gyr}$. Vertical and horizontal lines map out the quadrants defining star-forming stages (SB, SFMS, TZ, RS) and gas fraction domains (GP, GN, GR).}
\label{fig:tdep_two_panels}
\end{figure*}

To explore these mechanisms, we examine the distribution of $t_{\mathrm{dep}}$ in the $\Delta\text{SFMS}$--$\Delta\text{GFMS}$ plane shown in Fig.~\ref{fig:tdep_two_panels}. A critical threshold is highlighted at $t_{\mathrm{dep}} = 13.8\,\text{Gyr}$ (the Hubble time, $\log_{10}(t_{\mathrm{dep}}/\text{yr}) \sim 10.14$), which separates galaxies capable of exhausting their gas within a cosmic age from those whose star formation has slowed down so significantly that their depletion timescales stretch beyond the age of the Universe. A detailed census of the galaxies falling above and below this cosmic boundary across star-forming states and gas categories is presented in Table~\ref{tab:tdep_statistics}.

\begin{table*}
\centering
\caption{\textsc{Hi}-detection subpopulation breakdown and gas properties relative to the depletion boundary ($t_{\mathrm{dep}} \leq 13.8\,\text{Gyr}$ vs. $t_{\mathrm{dep}} > 13.8\,\text{Gyr}$) for both environments. The populations are divided by star-forming status (SB, SFMS, TZ, RS) and gas fraction offset categories: Gas Poor (GP; $\Delta\text{GFMS} < -0.3$), Gas Normal (GN; $-0.3 \leq \Delta\text{GFMS} \leq 0.3$), and Gas Rich (GR; $\Delta\text{GFMS} > 0.3$).} 
\label{tab:tdep_statistics}
\begin{tabular}{llccccccc}
\hline
\hline
\noalign{\smallskip}
 & & \multicolumn{3}{c}{\textbf{SSC cluster}} & & \multicolumn{3}{c}{\textbf{SSC filament}} \\
 \cline{3-5} \cline{7-9}
\noalign{\smallskip}
\textbf{Categorization} & \textbf{Type} & Total $N$ & $t_{\mathrm{dep}} \leq 13.8\,\text{Gyr}$ & $t_{\mathrm{dep}} > 13.8\,\text{Gyr}$ & & Total $N$ & $t_{\mathrm{dep}} \leq 13.8\,\text{Gyr}$ & $t_{\mathrm{dep}} > 13.8\,\text{Gyr}$ \\
\noalign{\smallskip}
\hline
\noalign{\smallskip}\\
\multicolumn{9}{l}{\textbf{\textup{Star-Forming States}}} \\
Starburst (SB)       & Dets     & 8   & 8  & 0   & & 30   & 28  & 2   \\ \\
Main Sequence (SFMS) & Dets     & 22  & 18 & 4   & & 62  & 49  & 13   \\ \\
Transition Zone (TZ) & Dets     & 93  & 50  & 43  & & 63  & 18  & 45  \\ \\
Red Sequence (RS)    & Dets     & 4  & 0  & 4  & & 2  & 0  & 2  \\
\noalign{\smallskip}
\hline
\noalign{\smallskip}\\
\multicolumn{9}{l}{\textbf{\textup{Gas Fraction Categories}}} \\
Gas Poor (GP)        & Dets     & 33  & 31   & 2  & & 9   & 8   & 1   \\ \\
Gas Normal (GN)      & Dets     & 77  & 40  & 37  & & 99  & 66   & 33  \\ \\
Gas Rich (GR)        & Dets     & 17   & 5   & 12   & & 49  & 21   & 28  \\
\noalign{\smallskip}
\hline \hline
\end{tabular}
\end{table*}

In the cluster centre (panel a), SB and SFMS detections predominantly exhibit short depletion timescales, with $8/8$ SB detections and $18/22$ ($81.8\,\mathrm{per\,cent}$) SFMS detections falling below the $13.8\,\mathrm{Gyr}$ threshold. As galaxies transition into the TZ, the distribution splits more evenly: out of 93 detected TZ galaxies in the cluster centre, 50 ($53.8\,\mathrm{per\,cent}$) maintain $t_{\mathrm{dep}} \le 13.8\,\mathrm{Gyr}$, while 43 ($46.2\,\mathrm{per\,cent}$) exceed the Hubble time. The cluster RS population exhibits complete quenching, with all 4 detected RS systems ($100\,\mathrm{per\,cent}$) featuring long depletion times ($t_{\mathrm{dep}} > 13.8\,\mathrm{Gyr}$).

Examining the cluster centre by gas subclasses reveals that GP detections ($31/33$, or $93.9\,\mathrm{per\,cent}$) predominantly reside at short depletion timescales, whereas GR detections skew toward longer depletion times ($12/17$, or $70.6\,\mathrm{per\,cent}$ exceeding $13.8\,\mathrm{Gyr}$). GN detections in the cluster centre show a balanced distribution, with 40 systems below and 37 above the Hubble time threshold.

This behavior differs from the SSC filament environment (panel b). Filament SB and SFMS populations show active gas consumption, with $28/30$ ($93.3\,\mathrm{per\,cent}$) of SB detections and $49/62$ ($79.0\,\mathrm{per\,cent}$) of SFMS detections exhibiting $t_{\mathrm{dep}} \le 13.8\,\mathrm{Gyr}$. However, as filament systems evolve into the TZ, long depletion times dominate: $45/63$ ($71.4\,\mathrm{per\,cent}$) of TZ detections exceed the $13.8\,\mathrm{Gyr}$ threshold. Similarly, the filament GR population shows a high fraction of extended depletion times ($28/49$, or $57.1\,\mathrm{per\,cent}$ with $t_{\mathrm{dep}} > 13.8\,\mathrm{Gyr}$), whereas the GN population is primarily concentrated at shorter depletion timescales ($66/99$, or $66.7\,\mathrm{per\,cent}$).

The filament RS population contains few detections ($2$ total), both of which exceed the Hubble time. These trends indicate that while filament TZ galaxies preserve substantial gas reservoirs, a significant fraction experience suppressed star formation efficiency—pointing to environmental stagnation or starvation within the filamentary network prior to cluster-centre infall.


\section{Discussion}
\label{sec:Discussion} 

This study presents an environmental analysis of the cold gas content, structural offsets, and consumption efficiencies of galaxies within the SSC complex, differentiating between the high-density cluster centre and the surrounding large-scale cosmic filaments. Below, we synthesize our empirical results to highlight the distinct physical mechanisms driving galaxy evolution across these cosmic structures.

\subsection{Environmental Imprints on the $\Delta\text{SFMS}$--$\Delta\text{GFMS}$ Plane}
\label{sec:disc_scaling_relations}

\subsubsection{The Cluster Centre: Accelerated Gas Deficits}

Within the high-density environment of the SSC centre, the distribution of galaxies across the star-forming and gas-fraction offset planes reveals a consistent, environmentally driven depletion sequence. As demonstrated in Fig.~\ref{fig:sSFR_gasF_scaling_relation}, our findings reveal that the cluster centre is dominated by GN, followed by GP systems that are concentrated within the TZ. This distribution reflects an evolutionary pathway where galaxies undergo both gas removal and star formation suppression as they migrate away from the main sequence. 

Cluster centre SB galaxies have a positive $\Delta\text{GFMS}$ offest of 0.08 dex, while galaxies residing on the active SFMS appear to maintain normal \textsc{Hi} gas reservoirs ($\Delta\text{GFMS} \sim -0.001\,\text{dex}$). However, once galaxies move below the SFMS into the TZ, their gas fractions decrease. The cluster TZ and RS populations exhibit progressively lower gas fraction offsets, with median values of $-0.13\,\text{dex}$ and $-0.16\,\text{dex}$, respectively (Table~\ref{tab:offset_statistics}). This systematic gas depletion in intermediate and quenched populations aligns with environmental quenching models \citep[e.g.][]{boselli2014cold,boselli2016quenching}, where ram-pressure stripping and starvation progressively remove gas reservoirs, driving galaxies off the SFMS toward the RS.  

This behaviour is consistent with the action of efficient environmental mechanisms operating within the virialized cluster environment. In particular, ram-pressure stripping (RPS) is expected to remove the outer, low-density \textsc{Hi} discs of galaxies as they move at high velocities through the dense intracluster medium (ICM), leading to rapid gas depletion and the suppression of star formation \citep{gunn1972infall,boselli2006environmental,cortese2021dawes, boselli2022ram,boselli2022virgo}. This interpretation is further supported by the small fraction of SB galaxies ($\sim 6$ per cent) among our cluster detections. Dense cluster environments are known to rapidly deplete the \textsc{Hi} gas reservoirs required to sustain enhanced star formation, thereby suppressing SB activity and accelerating the transition of galaxies towards the RS through environmental quenching \citep{larson1980evolution,boselli2016quenching,peng2015strangulation,reynolds2022wallaby}.

These observations are consistent not only with a pre-processing scenario, where galaxies begin to lose gas prior to cluster-centre infall, but also with environmental processes within the cluster itself, such as ram-pressure stripping and tidal interactions. The prolonged removal of gas—likely through starvation or strangulation—has been noted in other dense environments like the Coma and Virgo clusters \citep[e.g.,][]{yozin2015quenching, boselli2022virgo}, and is thought to drive the slow quenching pathways dominant in rich cluster environments.

\subsubsection{The Cosmic Filaments: Fuel Shielding}
Looking at the \textsc{Hi} detections in Fig.~\ref{fig:SFMS_GFMS_Staggered_Aligned} (panel d), a \textsc{Hi} gas excess spans all star-forming sub-classes uniformly, except the small and negligible RS population. Filamentary SB systems retain a gas excess ($\text{median} \sim 0.21\,\text{dex}$), with similarly high gas fractions in the SFMS ($\text{median} \sim 0.19\,\text{dex}$) and the TZ ($\text{median} \sim 0.15\,\text{dex}$). Our KS tests reveal no statistically significant differences between any of the evolutionary stages within the filaments (SB vs. SFMS: $p\approx0.44$; SFMS vs. TZ: $p\approx0.85$; SB vs. TZ: $p\approx0.12$), indicating that the gas fractions of detected galaxies remain stable from the SB to the SFMS. Given the \textsc{Hi}-selected nature of the sample, however, this result does not exclude differences in the underlying population that may become apparent with deeper observations.


\subsubsection{Cluster vs. Filament vs. xGASS Field}

To place the SSC results in the context of a lower-density field population, we compare the cluster and filament samples with the xGASS%
\footnote{\textcolor{blue}{\url{https://xgass.icrar.org/}}} survey \citep{catinella2018xgass,janowiecki2020xgass}. xGASS is a stellar-mass-selected, $M_\star > 10^9\ \text{M}_\odot$, gas-fraction-limited \textsc{Hi} survey of $\sim 1200$ galaxies with $0.02 < z < 0.05$, which is representative of the \textsc{Hi} properties of galaxies in the local Universe. Our SSC sample covers $0.036 < z < 0.058$ and extends to lower stellar masses, with $8.10 < \log_{10}(M_\star/\rm M_\odot) < 11.46$. The different stellar-mass and redshift selections, together with differences in observational sensitivity, mean that xGASS provides a reference population rather than a directly matched control sample. The xGASS results are therefore used to place the SSC measurements in the broader context of nearby field galaxies.


In the field, SFMS systems dominate the overall population, making up $52.3$ per cent ($614/1174$) of xGASS galaxies, while the global \textsc{Hi} detection rate reaches $64.0$ per cent ($751/1174$). The field TZ represents a long-lived, slow-evolving phase where galaxies gradually consume their cold gas via internal, secular evolution \citep[][]{janowiecki2020xgass}. When comparing our supercluster environments with this reference sample for the field, we observe:

\begin{enumerate}

    \item \textbf{dominance of processed populations:} the cluster centre shows a dominance of TZ galaxies among its secure \textsc{Hi} detections, constituting $73.2\,\text{per cent}$ ($93/127$). In the filaments, secure \textsc{Hi} detections are shared nearly equally between TZ systems ($40.1\,\text{per cent}$; $63/157$) and SFMS galaxies ($39.5\,\text{per cent}$; $62/157$). This significant concentration of TZ galaxies even within the filamentary network indicates that galaxy pre-processing takes place long before systems reach the cluster virial radii \citep[e.g.,][]{fujita2004pre,haines2015locuss,kraljic2018galaxy}.

    \item \textbf{lower incidence of detectable \textsc{Hi} gas:} the environmental trend is reflected in the observed \textsc{Hi} detection fractions (Fig.~\ref{fig:sample_overview_2x2}d). The observed \textsc{Hi} detection fraction is 45.0 per cent in the filaments and 30.0 per cent in the cluster centre, compared with 64.0 per cent in the xGASS reference sample. The higher detection fraction in xGASS provides useful context, but differences in stellar-mass range, redshift coverage, \textsc{Hi} selection, and observational sensitivity prevent a direct comparison of the intrinsic \textsc{Hi} detection fractions. We therefore interpret the cluster--filament comparison as the primary environmental comparison and use xGASS as a reference for the broader field population. The lower incidence of detectable \textsc{Hi} in the denser cluster environment is consistent with environmental processes that reduce or redistribute the atomic gas reservoirs of galaxies, including ram-pressure stripping, tidal interactions, and starvation \citep[e.g.][]{boselli2006environmental,hess2013evolution,cortese2021dawes,reynolds2022wallaby}.

    \item \textbf{depressed star formation levels:} In the field, GR and GN galaxies lie close to the SFMS ($\Delta\text{SFMS} \sim 0\,\text{dex}$), consistent with the well-established correlation between cold gas content and ongoing star formation in isolated galaxies \citep[e.g.][]{saintonge2011cold, catinella2018xgass, janowiecki2020xgass}. Among the \textsc{Hi}-detected galaxies in the cluster centre, all gas subclasses, including the GR population, exhibit negative median offsets below the SFMS, ranging from $\sim-0.79$ to $-0.43\,\text{dex}$ (Fig.~\ref{fig:SFMS_GFMS_Staggered_Aligned}, panel a). This indicates that an oversupply of \textsc{Hi} gas among detected galaxies does not guarantee active star formation in dense environments, where converting \textsc{Hi} gas into star-forming molecular clouds can be inefficient due to gas heating, turbulence, or dynamical perturbations \citep[e.g.][]{boselli2006environmental,Peng2010,boselli2014cold,boselli2016quenching}. In contrast, the \textsc{Hi}-detected galaxies in the filament environment (Fig.~\ref{fig:SFMS_GFMS_Staggered_Aligned}, panel b) show small negative median $\Delta\text{SFMS}$ offsets across all gas subclasses, with values remaining within the $\pm0.3\,\text{dex}$ scatter of the SFMS. This suggests that the detected filament population retains star formation activity close to the SFMS, although some galaxies show lower star formation rates. Overall, the median offsets indicate an intermediate level of star formation activity in the filaments compared with the field and cluster centre. This is broadly consistent with previous studies showing that filaments provide intermediate environments in which galaxy star formation can be enhanced or suppressed as galaxies approach cluster regions \citep[e.g.][]{salerno2019filaments,o2024effect}.
\end{enumerate}

In addition to exhibiting intermediate star formation activity between the field and cluster centre populations, the filamentary environment shows a distinct trend in \textsc{Hi} gas content. The \textsc{Hi}-detected filament galaxies exhibit higher gas fractions, with an average median, $\left<\Delta\text{GFMS}\right> \sim +0.12\,\text{dex}$, compared with both the field reference, $\left<\Delta\text{GFMS}\right> \sim -0.20\,\text{dex}$, and cluster centre, $\left<\Delta\text{GFMS}\right> \sim -0.05\,\text{dex}$, populations. This suggests that the filamentary structure represents an intermediate stage of environmental transformation, where galaxies retain a comparatively higher \textsc{Hi} gas content than those in the cluster centre, despite showing intermediate levels of star formation activity between the field and cluster populations.

The KS tests reveal two patterns when comparing the cluster centre and filament environments: \textbf{i)} star formation offsets depend on environment at fixed relative gas content---galaxies belonging to the same \textsc{Hi} gas class exhibit statistically distinct $\Delta\text{SFMS}$ distributions between the cluster centre and the filaments ($p \approx 0.04$ for GR; $p < 0.001$ for GN and GP). \textbf{ii)} \textsc{Hi} gas fraction distributions ($\Delta\text{GFMS}$) are statistically indistinguishable between the two environments across the SB ($p \approx 0.41$), SFMS ($p \approx 0.17$), and RS ($p \approx 0.93$) populations, differing only within the TZ ($p < 0.001$).

This indicates that the large-scale environment modulates the efficiency with which available \textsc{Hi} is converted into stars, rather than only setting the total gas reservoir. In particular, the divergence of $\Delta\text{GFMS}$ except in the TZ highlights this phase as the primary regime where environmental gas processing differentiates cluster-core galaxies from those in the surrounding filaments.


We stress that the strong environmental imprint found in the SSC may not characterize other dense environments. For example, \citet{davies2019dustpedia}, comparing Virgo galaxies with field galaxies, reported little to no difference in the interstellar medium scaling relations of galaxies, concluding that internal physical processes and morphology, rather than ongoing local environmental density, primarily dictate these properties. Studies of other nearby clusters and superclusters have nevertheless found clear environmental effects on \textsc{Hi} content, star formation, and galaxy properties, including in Virgo and the Coma Supercluster \citep[e.g.,][]{gavazzi2010snapshot,gavazzi2013h,boselli2014cold,cybulski2014voids,boselli2020virgo}. Filament studies around Virgo have also found evidence that galaxy properties vary between field, filament, and cluster environments, indicating that environmental effects can already be present within the filamentary network \citep[][]{castignani2022virgo, castignani2021virgo}.

This difference with the SSC highlights that while in relatively standard or more relaxed environments like Virgo the local density might not heavily alter the scaling relations, the extreme and unrelaxed dynamical conditions of a massive collapsing supercluster core like the SSC can override internal regulation, accelerating the environmental sequence and decoupling a galaxy's immediate gas content from its star formation output long before cluster-centre infall. It is important to note that the SSC is among the most massive and densest supercluster structures in the local Universe. The core of the SSC is an extraordinarily active dynamical region where cluster-cluster interactions and mergers are well-established by dynamical analyses, X-ray, and radio studies. Using the ShaSS data \citep{merluzzi2015shapley}, \citet{haines2018shapley} obtained a detailed map of the region around the supercluster core, which includes 11 SSC clusters. They showed that these 11 systems are all interconnected and lie within a coherent sheet of galaxies that fills the entire survey region without gaps \citep[see also][]{reisenegger2000shapley}. This study also revealed a prominent stream of galaxies connecting the northern cluster A3559 to the core.

\subsection{Disentangling Stellar Mass Dependencies and Environmental Variations}
\label{sec:mass_vs_environment}

\subsubsection{Internal Quenching Mechanisms Across Mass Ranges}

When the supercluster population is separated into different masses, low-- ($M_{\star} < 10^{10}\,\text{M}_{\odot}$) and high--mass ($M_{\star} \geq 10^{10}\,\text{M}_{\odot}$) bins, a mass-dependent signature emerges within the $\Delta\text{SFMS}$ distributions. As shown in the top row of Fig.~\ref{fig:SFMS_GFMS_histograms} (panels a--c), moving from low-mass to high-mass subsamples systematically drives the $\Delta\text{SFMS}$ into negative values. This trend is observed across both environments, with the median offset of \textsc{Hi} detections falling from $-0.45\,\text{dex}$ to $-0.67\,\text{dex}$ in the cluster centre, and from $-0.15\,\text{dex}$ to $-0.20\,\text{dex}$ in the filaments. 

The physical driver behind this downward trend is the action of mass-dependent internal quenching mechanisms—such as active galactic nucleus (AGN) feedback or morphological stabilization—which scale strongly with the depth of the deep potential well in massive galaxies \citep{martig2009morphological}. In the high-mass range ($M_{\star} \geq 10^{10}\,\text{M}_{\odot}$; panel c), the integrated mean of medians (marked by the crosses) drops to roughly $-0.94\,\text{dex}$ and $-0.36\,\text{dex}$ for the cluster and filament, respectively. This shows that regardless of the large-scale environment, massive systems are heavily quenched, confirming that internal, mass-governed processes remain efficient at suppressing star formation even within large-scale cosmic web structures \citep{Peng2010}. 

From a theoretical point of view, hydrodynamical studies such as the IllustrisTNG simulations provide valuable context for these empirical trends. For instance, \citet{donnari2021quenched} demonstrate that high-mass galaxies, $M_{\star} \geq 10^{10.0-10.5}\,\text{M}_{\odot}$, exhibit an intrinsic tendency to halt star formation via internal AGN feedback channels independently of their broader cosmic surroundings. This picture is further corroborated by \citet{goubert2024role} and \citet{hasan2024filaments}, who argue that internal galaxy parameters—most notably central black hole mass—serve as the dominant statistical predictors of quiescence within central systems and high-mass satellites. 


\subsubsection{Cross-Environmental Gas Divergence}

While the $\Delta\text{SFMS}$ distributions are dominated by mass-driven signatures, the $\Delta\text{GFMS}$ detailed in the bottom row of Fig.~\ref{fig:SFMS_GFMS_histograms} reveal a significant divergence driven by environment. This cross-environmental separation is confirmed by KS tests, which demonstrate that the underlying gas fraction distributions are statistically distinct across both mass bins for the \textsc{Hi} detections (low mass $p \approx 0.02$, high mass $p < 0.001$).

In the cluster environment, the overall sample of detections sits near the nominal gas fraction ($\text{median}\ \Delta\text{GFMS} = -0.02\,\text{dex}$; Fig.~\ref{fig:SFMS_GFMS_histograms}d). Splitting by stellar mass shows that low-mass sample retain a slight gas fraction excess ($\text{median}\ \Delta\text{GFMS} = +0.06\,\text{dex}$; panel e), whereas high-mass sample exhibit a gas fraction deficit ($\text{median}\ \Delta\text{GFMS} = -0.24\,\text{dex}$; panel f). 

By comparison, the filament environment detections show an high \textsc{Hi} gas fraction across both mass bins (Fig.~\ref{fig:SFMS_GFMS_histograms}e, f). Filament detections show positive median gas fraction offsets for low-mass systems ($+0.26$ dex), while high-mass systems have a median offset of $-0.03$ dex. Consequently, across both stellar mass regimes, cluster galaxies systematically exhibit smaller $\Delta\text{GFMS}$ than their filament counterparts—a shift of $0.20\,\text{dex}$ lower in low-mass systems and $0.21\,\text{dex}$ lower in high-mass systems. This indicates that \textsc{Hi} gas reduction is actively underway in cluster galaxies, even though these detected systems are not yet totally GP. 
Similar to findings in the Hydra~I cluster \citep{reynolds2022wallaby}, these detections are consistent with an earlier stage of environmental \textsc{Hi} gas loss compared to the more strongly depleted populations observed in evolved clusters such as Virgo \citep[e.g. VIVA;][]{chung2009vla,boselli2014cold}. Similar environmental trends have also been reported in other nearby clusters and supercluster environments \citep[][]{gavazzi2010snapshot,gavazzi2013h,cybulski2014voids,boselli2020virgo}.

\subsection{Environmental Modulations of \textsc{Hi} Depletion Timescales}
\label{sec:disc_tdep}

The mapping of gas depletion timescales ($t_{\mathrm{dep}}$) for secure \textsc{Hi} detections within the joint $\Delta\text{SFMS}$--$\Delta\text{GFMS}$ space (Fig.~\ref{fig:tdep_two_panels}) provides two distinct quenching pathways across the supercluster, tightly coupled to the morphological transformation of the galaxy population.

In the high-density cluster centre, the transition of $\textsc{Hi}$ detections into the extended depletion regime ($t_{\mathrm{dep}} > 13.8\,\text{Gyr}$) is synchronized with severe gas loss. Among secure detections entering the TZ, the distribution splits nearly evenly, with $53.8\,\text{per cent}$ ($50/93$) maintaining $t_{\mathrm{dep}} \le 13.8\,\text{Gyr}$ while $46.2\,\text{per cent}$ ($43/93$) exceed the Hubble time. This points to a progressive suppression of star formation rates relative to remaining gas reservoirs. Within the cluster centre, secure \textsc{Hi} detections span diverse gas reservoirs, with the GP subset ($N=33$) reflecting environments where ram-pressure stripping efficiently shears away outer, low-density \textsc{Hi} gas disks on short timescales, while compact central molecular reservoirs temporarily continue to fuel star formation \citep{bahe2015star, boselli2022ram}. This gas removal and depletion drives the morphological transformation of the cluster centre, as seen in Table~\ref{tab:morphological}, where the vast majority of the passive, non-detected cluster population has been converted into ETGs ($88\,\text{per cent}$).

Conversely, the secure \textsc{Hi} detections in the filaments follow a different evolutionary path. Rather than sliding down a simple diagonal gas-loss track, filament detections exhibit a significant accumulation of long depletion times as they migrate off the main sequence: $71.4\,\text{per cent}$ ($45/63$) of filament TZ detections and $57.1\,\text{per cent}$ ($28/49$) of filament GR detections exceed $t_{\mathrm{dep}} = 13.8\,\text{Gyr}$. This combination of preserved gas fraction offsets and inefficient star formation suggests that the filamentary environment allows galaxies to retain substantial \textsc{Hi} reservoirs, either because they have not yet experienced the hydrodynamic stripping operating in the cluster centre or because they continue to acquire gas through diffuse accretion. Despite this, the conversion of $\textsc{Hi}$ gas into stars remains inefficient \citep{kleiner2017evidence,vulcani2019gasp}.

A crucial insight into the broader evolutionary history of the filaments emerges when aligning these depletion tracks with the morphological census from Table~\ref{tab:morphological}. While the secure \textsc{Hi} filament detections are dominated by star-forming morphologies (LTGs at $72\,\text{per cent}$ and Irregulars at $26\,\text{per cent}$), the underlying non-detected filament population is dominated by ETGs ($73\,\text{per cent}$). This morphological difference provides evidence that environmental pre-processing is already active within the filamentary structures \citep[e.g.,][]{fujita2004pre,cybulski2014voids,bianconi2018locuss}.  

The long $t_{\mathrm{dep}}$ ($\sim 40$ to $>100\,\text{Gyr}$) in these gas-rich filament systems point to a global mechanism of star formation stagnation or starvation \citep{peng2015strangulation, van2017environmental}. These extended depletion timescales reflect conditions where local disk surface densities and ambient thermal pressures fall below thresholds necessary to induce rapid gravitational collapse \citep{leroy2008star, bigiel2011constant, schruba2011molecular}. 

While molecular gas ($\text{H}_2$) maintains a direct link to active star-forming sites \citep[e.g.,][]{leroy2008star}, large-scale atomic reservoirs can become dynamically stable and inefficient at forming stars if they are prevented from cooling and collapsing into dense molecular clouds \citep[e.g.,][]{catinella2018xgass,janowiecki2020xgass}. Spatially resolved studies of nearby spirals show that while star formation remains locally governed by the molecular component, $\textsc{Hi}$ decorrelates from individual star-forming sites, reinforcing its role as a diffuse structural reservoir rather than an immediate tracer of ongoing star formation \citep[e.g.,][]{bigiel2008star,casasola2022resolved}. Within the filaments, the ambient environment may maintain or replenish large-scale diffuse $\textsc{Hi}$ envelopes while lacking the peak hydrodynamic pressures required to rapidly remove this gas before it reaches denser regions \citep[e.g.,][]{boselli2006environmental,cortese2021dawes}. Consequently, these $\textsc{Hi}$-detected filament TZ and/or GR galaxies are consistent with a form of pre-processing in which star formation is suppressed while substantial $\textsc{Hi}$ reservoirs are retained.

The high fraction of quenched, non-detected ETGs in the filamentary structure suggests that this pre-processing pathway has already affected a major fraction of the filament population. Active star-forming disks can undergo substantial transformation into inactive galaxies before entering the virialised cluster centre, consistent with studies showing that galaxy evolution begins in groups and filaments prior to cluster infall \citep[e.g.,][]{fujita2004pre,kraljic2018galaxy,o2024effect}. The surviving gas-rich LTG and irregular detections with $t_{\mathrm{dep}} \gg 13.8\,\text{Gyr}$ thus capture the final, visible phase of this transition, acting as an evolutionary bridge between isolated field spirals and fully processed cluster spheroids.

\section{Summary and Conclusions}
\label{sec:Summary_Conclusions}

In this paper, we investigate the \textsc{Hi} properties of galaxies across the Shapley Supercluster (SSC) by combining new deep MeerKAT L-band observations of five filament pointings with archival MeerKAT Galaxy Cluster Legacy Survey (MGCLS) observations of the dense supercluster core \citep{knowles2022meerkat}. Together, these data provide a homogeneous view of the \textsc{Hi} gas content of galaxies spanning environments from the virialised cluster centre to the large-scale filaments connecting the SSC--core to Abell\,3559 and Abell\,3560. By combining the \textsc{Hi} data with the extensive redshift coverage of the SSC provided by the Shapley Supercluster Survey project \citep[ShaSS,][]{merluzzi2015shapley}, we investigate how environment influences the gas content, star formation activity, and evolutionary state of \textsc{Hi}-detected galaxies. Our analysis includes the spatial distribution of galaxies, \textsc{Hi} scaling relations ($\Delta\mathrm{GFMS}$), star formation offsets ($\Delta\mathrm{SFMS}$), gas depletion timescales ($t_{\rm dep}$), and morphological properties. The analysis is based on samples comprising 127 \textsc{Hi}-detected cluster galaxies and 157 \textsc{Hi}-detected filament galaxies, within the ShaSS redshift range, $0.036 < z < 0.058$. The cluster observations reach an \textsc{Hi} mass detection limit of $M_{\textsc{Hi}} \sim 3.35 \times 10^{8}\ \textup{M}_{\odot}$, corresponding to a column density sensitivity of $N_{\textsc{Hi}} \sim 4.94 \times 10^{19}\ \textup{atoms\ cm}^{-2}$ \citep[][]{gwebushe2026neutral}. The filament observations have an \textsc{Hi} mass detection limit of $M_{\textsc{Hi}} \sim 3.11 \times 10^{8}\,\mathrm{M}_{\odot}$ and a corresponding column density sensitivity of $N_{\textsc{Hi}} \sim 4.59 \times 10^{19}\,\mathrm{atoms\ cm}^{-2}$. Our main results can be summarised as follows: 

\begin{enumerate}

\item Combined with the previously analysed MGCLS observations, our new MeerKAT data provide one of the largest contiguous \textsc{Hi} surveys of a nearby supercluster, allowing a direct comparison between galaxies residing in the dense cluster centre and those located within the surrounding filamentary region.

\item The \textsc{Hi} detection fraction decreases with increasing environmental density within the SSC. The filament population exhibits a detection fraction of approximately 45 per cent, while $\sim$30 per cent of galaxies are detected within the virialised cluster centre. Given the comparable sensitivity of the cluster and filament observations, this difference indicates a lower incidence of detectable \textsc{Hi} in the denser cluster environment, consistent with environmental processing of the \textsc{Hi} gas.

\item The cluster population of \textsc{Hi} detections is dominated by transition-zone systems, whereas the filament population is shared nearly equally between transition-zone and star-forming main sequence galaxies, indicating that galaxies across the supercluster are, on average, more advanced along the quenching sequence than typical field populations.


\item The cluster centre and filamentary structures exhibit distinct evolutionary signatures. Cluster galaxies progressively become more \textsc{Hi}-deficient as they evolve from the star-forming main sequence towards the red sequence, consistent with efficient environmental gas removal. In contrast, many filament galaxies retain relatively large atomic gas reservoirs. 

\item The \textsc{Hi}-detected gas-rich filament galaxies possess gas depletion times well in excess of the Hubble time ($t_{\text{dep}}\gg 13.8\ \text{Gyr}$), indicating that their \textsc{Hi} gas reservoirs are preserved but are converted into stars inefficiently. These results support a scenario in which galaxies begin to experience environmental pre-processing within the filament before entering the virialised cluster environment.


\end{enumerate}

Overall, our results indicate that galaxy evolution across the SSC is governed by multiple environmental mechanisms acting over different spatial scales. Within the dense cluster centre, galaxies experience efficient depletion of their \textsc{Hi} gas reservoirs, leading to the suppression of star formation and the build-up of the passive population. In contrast, while active star-forming galaxies in the filaments maintain efficient depletion timescales, systems migrating into the transition zone and gas-rich populations preserve substantial \textsc{Hi} reservoirs while exhibiting low star formation efficiencies and long gas depletion times. This behaviour is consistent with environmental pre-processing through starvation or strangulation, where the supply of fresh gas for star formation is reduced before galaxies are accreted into the cluster environment. This interpretation is further supported by the morphological distributions: while the \textsc{Hi}-detected filament population is dominated by late-type galaxies, the overall filament population already contains a substantial fraction of early-type galaxies, indicating that morphological transformation is underway before galaxies enter the cluster centre. Together, these results indicate that environmental transformation begins well beyond the virial radius and that the filamentary structures of the SSC play an important role in shaping the evolution of galaxies.

This work demonstrates the power of MeerKAT in tracing the influence of large-scale structure on galaxy evolution, particularly when combined with rich ancillary datasets like ShaSS. Moreover, the SSC represents an excellent target for future observations with MeerKAT$+$ and the SKA, promising even deeper insights into environmental effects on galaxy evolution.

\section*{Acknowledgements}
We thank X. Ndaliso, K. Trehaeven, E. Elson, S. Makhathini for helpful discussions. The authors also thank C. P. Haines and C. Tortora for valuable discussions. The MeerKAT telescope is operated by the South African Radio Astronomy Observatory, which is a facility of the National Research Foundation, an agency of the Department of Science and Innovation. The financial assistance of the South African Radio Astronomy Observatory (SARAO%
\footnote{\textcolor{blue}{\url{https://www.sarao.ac.za/}}}) towards this research is hereby acknowledged. This work was supported in part by the Italian Ministry of Foreign Affairs and International Cooperation, grant number ZA23GR03. LG acknowledge funding from the South African Radio Astronomy Observatory, which is a facility of the National Research Foundation (NRF), an agency of the Department of Science and Innovation (DSI). LG, TV, GB, PM, and VC acknowledge support from the INAF Mini-Grant 2024, "\textsc{Hi}--GalESS Tracing galaxy evolution in the Shapley Supercluster with \textsc{Hi}". The research of OS is supported by the South African Research Chairs Initiative of the DSTI/NRF (grant No. 81737). The authors also acknowledge the use of the Ilifu%
\footnote{\textcolor{blue}{\url{https://www.ilifu.ac.za/}}} cloud computing facility, a partnership between the University of Cape Town, the University of the Western Cape, the University of Stellenbosch, Sol Plaatje University, the Cape Peninsula University of Technology and the South African Radio Astronomy Observatory. The Ilifu facility is supported by contributions from the Inter-University Institute for Data Intensive Astronomy (IDIA; a partnership between the University of Cape Town, the University of Pretoria, the University of the Western Cape and the South African Radio Astronomy Observatory), the Computational Biology division at UCT, and the Data Intensive Research Initiative of South Africa (DIRISA). This research made use of Astropy%
\footnote{\textcolor{blue}{\url{https://www.astropy.org/}}}, a community-developed core Python package for astronomy. The authors thank the anonymous referee for her/his constructive comments and suggestions. The authors also thank the Scientific Editor for their comments and suggestions.  
\section*{Data Availability}
The raw MeerKAT data underlying this article is available on
the SARAO archive, at \textcolor{blue}{https://archive.sarao.ac.za}. The MeerKAT reduced cubes, derived data products, and/or galaxy properties used in this work are available upon reasonable request from the corresponding authors. 



\bibliographystyle{mnras}
\bibliography{example} 








\bsp	
\label{lastpage}
\end{document}